\documentclass{article}

\usepackage{PRIMEarxiv}

\usepackage[utf8]{inputenc} 
\usepackage[T1]{fontenc}    
\usepackage{hyperref}       
\usepackage{url}            
\usepackage{booktabs}       
\usepackage{amsfonts}       
\usepackage{nicefrac}       
\usepackage{microtype}      
\usepackage{lipsum}
\usepackage{fancyhdr}       
\usepackage{graphicx}       
\graphicspath{{media/}}     
\usepackage{authblk}
\usepackage{todonotes}
\usepackage{xspace}
\usepackage{float}
\usepackage{subcaption}
\usepackage[nolist]{acronym}
\usepackage{amsmath,amssymb,amsfonts}
\usepackage{booktabs}%
\usepackage{multirow}%

\newcommand{\arch}{DeepRed\xspace}

\title{DeepRed: an architecture for redshift estimation}

\author[1,*]{Alessandro Meroni}
\author[1,2,*]{Nicolò Oreste Pinciroli Vago}
\author[1]{Piero Fraternali}

\affil[1]{\small Department of Electronics, Information and Bioengineering, Politecnico di Milano, Milan, Italy}
\affil[2]{\small INAF -- Osservatorio Astronomico di Roma, Monteporzio Catone, RM, Italy}
\affil[*]{\small Equal contribution}

\affil[ ]{\small \texttt{\{alessandro5.meroni, nicolooreste.pinciroli, piero.fraternali\}@polimi.it}}

\begin{document}
\maketitle

\begin{acronym}
\acro{AI}{Artificial Intelligence}
  \acro{HOG}{Histogram of Oriented Gradients}
  \acro{SVM}{Support Vector Machine}
  \acro{SVR}{Support Vector Regressor}
  \acro{NLP}{Natural Language Processing}
  \acro{ViT}{Vision Transformers}
  \acro{CNN}{Convolutional Neural Network}
  \acro{SHAP}{SHapley Additive exPlanations}
  \acro{MSE}{Mean Square Error}
  \acro{MAE}{Mean Absolute Error}
  \acro{PCA}{Principal Component Analysis}
  \acro{MLP}{Multi-Layer Perceptron}
  \acro{ResNet}{Residual Network}
  \acro{ReLU}{Rectified Linear Unit}
  \acro{ML}{Machine Learning}
  \acro{DL}{Deep Learning}
  \acro{ANN}{Artificial Neural Network}
  \acro{RBF}{radial basis function}
  \acro{CC}{core-collapsed}
\acro{RNN}{Recurrent Neural Network}
\acro{XAI}{Explainable Artificial Intelligence}
\acro{LRP}{Layer-wise Relevance Propagation}
\acro{LSST}{Large Synoptic Survey Telescope}
\acro{ESO}{European Southern Observatory}
\acro{LIME}{Local Interpretable Model-Agnostic Explanations}
\acro{FCN}{Fully Connected Network}
\acro{MDN}{Mixture Density Network}
\acro{LSNIa}{Lensed Supernova of type Ia}
\acro{LSNcc}{Lensed Supernova of type core-collapse}
\acro{DES}{Dark Energy Survey}
\acro{CV}{Computer Vision}
\acro{LR}{Learning Rate}
\acro{SIFT}{Scale-Invariant Feature Transform}
\acro{ResNet}{Residual Network}
\acro{FC}{Fully Connected}
\acro{CAM}{Class Activation Map}
\acro{GT}{Ground Truth}
\acro{NN}{Neural Network}
\acro{LRE}{Linear Regression Ensemble}
\acro{KiDS}{Kilo-Degree Survey}
\acro{SDSS}{Sloan Digital Sky Survey}
\acro{MLPm}{MLP-mixer}
\acro{SwinT}{Swin Transformer}
\acro{DGL}{DeepGraviLens}
\acro{NMAD}{Normalized Mean Absolute Deviation}
\acro{PSF}{Point Spread Function}
\end{acronym}

\begin{abstract}
Estimating redshift is a central task in astrophysics, but its measurement is costly and time-consuming. In addition, current image-based methods are often validated on  homogeneous datasets. The development and comparison of networks able generalize across different morphologies, ranging from galaxies to gravitationally-lensed transients, and  observational conditions,  remain an open challenge. This work proposes \arch, a deep learning pipeline that demonstrates how modern computer vision architectures, including ResNet, EfficientNet, Swin Transformer, and MLP-Mixer, can estimate redshifts from images of galaxies, gravitational lenses, and gravitationally-lensed supernovae. We compare these architectures and their ensemble to both neural networks (A1, A3, NetZ, and PhotoZ) and a feature-based method (HOG+SVR) on simulated (DeepGraviLens) and real (KiDS, SDSS) datasets. Our approach achieves state-of-the-art results on all datasets. On DeepGraviLens, \arch achieves a significant improvement in the Normalized Mean Absolute Deviation (NMAD) compared to the best baseline (PhotoZ): 55\% on DES-deep (using EfficientNet), 51\% on DES-wide (Ensemble), 52\% on DESI-DOT (Ensemble), and 46\% on LSST-wide (Ensemble). On real observations from the KiDS survey, the pipeline outperforms the best baseline (NetZ), improving NMAD by 16\% on a general test set without high-probability lenses (Ensemble) and 27\% on high-probability lenses (Ensemble). For non-lensed galaxies in the SDSS dataset, the MLP-Mixer architecture achieves a 5\% improvement over the best baselines (A3 and NetZ). In addition, SHapley Additive exPlanations (SHAP) shows that the models correctly focus on the objects of interest with over 95\% localization accuracy on high-quality images, validating the reliability of the predictions. These findings suggest that deep learning architectures are a scalable, robust, and interpretable solution for redshift estimation in future large-scale surveys comprising heterogeneous astronomical objects.
\end{abstract}


\section{Introduction}
Astrophysics is the branch of astronomy that employs the principles of physics and chemistry to ascertain the nature of astronomical objects \cite{Carroll2017}. Astrophysics has led to the discovery of new phenomena, such as gravitational lenses \cite{Einstein1936}, spatio-temporal distortions caused by the gravitational field of massive objects. When an astronomical source is behind a gravitational lens relative to an observer, it appears distorted, often forming structures like Einstein rings \cite{jauncey_unusually_1991} and Einstein crosses \cite{turyshev_wave-optical_2021}.

Time-domain astronomy \cite{bernardini_astronomy_2011} studies transient astronomical events (i.e., short-lived events with durations ranging from fractions of a second to years), such as supernovae \cite{burrows_core-collapse_2021} and pulsars \cite{ambrosino_optical_2021, israel_beyond_2025, pintore_new_2025, rea_fifty_2017, reardon_bow_2025}. Supernovae are powerful and luminous stellar explosions that occur during the last evolutionary stages of a massive star. Gravitational lenses can distort supernovae, giving origin to gravitationally-lensed supernovae \cite{kodiramanah_ai-driven_2022, patel_three_2014, rodney_gravitationally_2021}, a rare phenomenon that allows cosmological studies and the testing of relativity.

Redshift is the shift of an object's spectral lines toward longer wavelengths (i.e., toward red in the visible spectrum) compared to a reference. This phenomenon can occur due to the universe's expansion (cosmological redshift \cite{mamas_explanation_2010}), the relative motion of the object with respect to the observer (Doppler redshift \cite{bedran_comparison_2002, bunn_kinematic_2009}), or strong gravitational fields (gravitational redshift \cite{wilhelm_gravitational_2014}). Redshift can be estimated either through spectroscopy (spectroscopic redshift, based on the analysis of objects' spectra) or photometry (photometric redshift, based on measurements of objects' brightness across different frequency ranges). Redshift has several applications in astrophysics, including estimating distances between astronomical objects \cite{distance}, creating 3D maps of the universe \cite{York_2000}, and estimating the universe expansion rate \cite{Riess1998}.

Modern astrophysics benefits from large amounts of data \cite{kremer_big_2017, sen_astronomical_2022} acquired by ground-based and space telescopes across multiple wavelengths (e.g., in the optical and X-ray bands). Large-scale surveys, such as \ac{LSST} \cite{Schwamb2023}, are expected to generate petabytes of data \cite{LSST_big_data}. Handling large amounts of data presents significant challenges, including data storage, processing, and analysis, making manual analysis impractical. For this reason, data-driven techniques are necessary. 

\ac{AI} can support data-driven tasks in astrophysics, as it allows the analysis of large amounts of data, including the classification of objects \cite{PinciroliVago2023} and their automated detection. The work in \cite{Baqui2021} uses \ac{ML} to classify stars, the work in \cite{Martens2011} uses computer vision to study solar activity, and the work in \cite{Valizadegan2022} uses deep learning to detect and classify exoplanet candidates with high precision automatically. The works in \cite{Henghes2022, schuldt_photometric_2021} focus on redshift estimation and highlight the challenges in obtaining spectroscopic redshifts, which require significant telescope time and human resources. The work in \cite{Li2021} estimates redshifts for quasars using \ac{ML}, and the work in \cite{Chunduri} proposes \acp{CNN} and \ac{DL} for redshift estimation. Ensuring the trustworthiness of \ac{AI} models remains a significant challenge in astrophysics, as not all the existing \ac{DL} algorithms used in this field are interpretable. The survey in \cite{lieu_comprehensive_2025} presents an overview of interpretability methods for \ac{AI} in astronomy.

The primary goal of this work is to estimate redshift for gravitational lens sources and galaxies using \ac{CV} algorithms. For this purpose, we use two real datasets and four simulated labelled datasets. 

The four simulated datasets contain $\approx15,000$ gravitational lenses each and are extracted from the \ac{DGL} dataset \cite{zenodo, PinciroliVago2023, pinciroli_vago_multimodal_2025}. These datasets replicate the observations captured with LSSTCam \cite{Stalder2020} or DECam \cite{Flaugher_2015}. The \ac{KiDS} real dataset comprises $\approx4,500$ candidate gravitational lenses, classified in \cite{Grespan2024} and retrieved from the \ac{ESO} Science Archive Facility. Each image is paired with a photometric redshift estimated using more spectral channels than those available in the image itself. This dataset serves as an additional validation of the model's performance in a real scenario. The \ac{SDSS} real dataset comprises $\approx517,000$ galaxies, paired with spectroscopic redshifts. We intentionally consider heterogeneous regimes to test the limits of image-only redshift estimators. We apply the same pipeline to different morphologies and data (simulated vs. real) to  demonstrate that modern computer vision backbones can estimate redshift independently of specific object types or \acp{PSF}.

The contributions of this work can be summarized as follows:

\begin{itemize}
    \item We introduce the architecture of \arch, a pipeline that takes as input images of astronomical observations and produces as output redshift estimates. \arch uses complementary sub-networks trained independently and ensembles their outputs by means of a \ac{LRE}.
    \item We evaluate the designed architecture on the four simulated \ac{DGL} datasets \cite{zenodo}, on the real \ac{KiDS} dataset and on the real \ac{SDSS} dataset.
    \item We compare \arch with \ac{HOG} + \ac{SVR}, A1 \cite{architecture}, A3 \cite{architecture}, NetZ \cite{schuldt_photometric_2021} and PhotoZ \cite{pasquet_photometric_2019} using a set of established metrics in redshift estimation (\ac{NMAD}, $\sigma_{68}$, bias and outlier rate). \arch yields \ac{NMAD} improvements ranging from $\approx 46\%$ to $\approx 55\%$ on the simulated datasets and from $\approx 5\%$ to $\approx 29\%$ on the real datasets. 
    \item We use \ac{SHAP}, an explainability algorithm, to show that \arch focuses on the relevant parts of the images (i.e., the astronomical objects) rather than on irrelevant areas of the background. To quantify the explainability results, we extract the most influential pixel for each image and compute the localization accuracy with respect to the astronomical object bounding box (over $95\%$ on datasets with less noise).
    \item We demonstrate that \arch generalizes well across datasets with different observational conditions, showing its suitability for future large-scale sky surveys.
\end{itemize}

To summarize, \arch is a methodological framework and ensemble pipeline, rather than a single new neural network architecture. While the individual components rely on established backbones, the primary contribution is the integration of distinct architectural families, to exploit their differences. 
Furthermore, \arch incorporates explainable AI not only as a post-hoc analysis, but as a validation tool that ensures that the redshift predictions are derived from the most relevant parts of the images.


The rest of the paper is organized as follows: Section \ref{sec:related} presents an overview of the state of the art and the astrophysics background concepts; Section \ref{sec:dsmeth} presents the dataset, the metrics, and the algorithms used in this work; Section \ref{sec:results} presents quantitative results and explainability results; Section \ref{sec:conclu} presents the conclusions and discusses potential future directions for research based on this work.

\section{Related work}
\label{sec:related}

\ac{ML} is a branch of \ac{AI} that focuses on systems that can learn from data. \ac{DL} is a branch of \ac{ML} that uses \acp{ANN} formed by multiple layers and allows to learn complex patterns in data. It can handle different types of data  \cite{Sarker2021}, including tabular data, images, and sequential data (e.g., time series and videos).

The work in \cite{Sarker2021} identifies three main architecture categories: \ac{MLP} (for tabular data), \ac{CNN} (for images), and  \ac{RNN} (for sequential data). Since 2017, transformers \cite{attention} have been used to address \ac{NLP} tasks and are now also used for vision tasks. \ac{ViT} \cite{VIT}  model images as sequences of patches, each enriched with positional information. They apply self-attention to all patch pairs, focusing on the most relevant regions and capturing global dependencies across the image. Compared to \acp{CNN}, transformers are better at modeling long-range relationships. Variants like \ac{SwinT} \cite{swin} incorporate hierarchical designs and local attention to balance detailed and generic understanding of the images.

\ac{CV} is an interdisciplinary field concerned with the automatic extraction of information and patterns from images \cite{computer_vision}. It has advanced over the years, driven by improvements in \ac{DL} architectures. Traditional methods such as  \ac{SIFT} \cite{sift}, \ac{HOG} \cite{HOG}, and edge detection algorithms like Canny \cite{Canny1986} extract features that can be combined with classifiers like \ac{SVM} or regressors like \ac{SVR} \cite{Comito2022}. These methods are effective for datasets with fewer data but are less effective than \acp{CNN} on bigger datasets  \cite{Sarker2021}. \ac{DL} revolutionized computer vision, becoming common for solving modern vision tasks. \acs{ResNet}-based \cite{resnet} and transformer-based architectures, for instance, are effective at object detection and segmentation.

Image regression is a type of regression task in \ac{CV}, where the input is an image and the output consists of one or more continuous numerical variables, rather than class labels. Image regression has several applications, such as medicine \cite{miao_cnn_2016} and environmental monitoring \cite{sarkar_real-time_2024}. The work in  \cite{hoekendijk_counting_2021} addresses the task of counting objects in ecological surveys, and the work in \cite{pinciroli_vago_multimodal_2025} extends image regression to the case of multimodal astronomical data.

Ensembling is an approach to improve prediction performances starting from the predictions of multiple \ac{ML} models. The survey in \cite{Mienye2022} presents diverse ensembling applications, ranging from medical diagnosis to fraud detection and astrophysics. The work in \cite{Wang2019} uses ensembling for pulsar candidate classification, and the work in \cite{PinciroliVago2023} uses ensembling for gravitationally-lensed supernovae classification. Ensembling methods, used for both classification and regression, include key techniques like stacking, boosting, and bagging \cite{Mienye2022}. 

\ac{DL} architectures are often considered black-box models due to their complexity, making their decision processes hard to interpret. \ac{XAI} addresses this challenge by enabling the interpretation of \ac{DL} outputs, with several key goals: to justify the model's outputs (\textit{explain to justify}), assess and monitor decisions to detect vulnerabilities or biases (\textit{explain to control}), identify areas for performance improvement (\textit{explain to improve}), and uncover new patterns and insights (\textit{explain to discover}) \cite{A2023}. The work in \cite{A2023} presents several \ac{XAI} techniques. Grad-CAM \cite{gradcam} is a variant of \ac{CAM} that creates heatmaps highlighting important regions of the input image, using gradients from the target class to compute the importance of each region. Still, the most effective \ac{CAM} variant depends on the specific problem \cite{gradcam,milani_proposals_2022, na_improving_2024, pinciroli_vago_comparing_2021}. \ac{LIME} \cite{lime} uses simple surrogate models to approximate the predictions of the underlying model. \ac{LRP} \cite{lrp} creates a heatmap of the input by working backwards from the output, starting from the last layer. Finally, \ac{SHAP} \cite{SHAP} is a game-theoretic approach for explaining machine learning model outputs by perturbing input data and measuring how much the predictions change. Compared with \ac{LRP} and \ac{CAM}-based algorithms, \ac{SHAP} is model-agnostic, making it a more versatile approach. Compared with \ac{LIME}, \ac{SHAP} is regarded as easier to interpret in \cite{salih_perspective_2025}.

Time-domain astronomy studies transient astronomical events, such as 
supernovae (i.e., powerful and luminous explosions of stars occurring during the last evolutionary stages of massive stars), pulsars (i.e., highly magnetized rotating neutron stars that emit beams of electromagnetic radiation out of their magnetic poles) \cite{HEWISH1968}, and gravitationally-lensed transients (i.e., the spatio-temporal distortion caused by massive objects such as galaxies and black holes lensing transient phenomena).

One key aspect of understanding these phenomena is determining their redshift, which provides information about their distance and the expansion of the universe. Redshift can be estimated using spectroscopic and photometric methods \cite{FernandezSoto2001}. The spectroscopic redshift \cite{spectroscopic} uses emission spectra (i.e., the light emitted over a range of energies), which are determined by the chemical composition of astronomical bodies. In the case of redshift, the spectra are shifted with respect to theoretical predictions, making it measurable. This method is time-consuming and requires high-quality and expensive spectroscopes for accurate measurements. Photometric redshift \cite{photometric} compares data of new and already known astronomical objects using brightness values for specific energy bands. This method is faster and cheaper than spectroscopic redshift, but also less accurate.

\ac{ML} has demonstrated its effectiveness in estimating both photometric and spectroscopic redshifts from images. The work in \cite{Sadeh2016} applies \ac{ML} to photometric measurements, specifically the magnitudes of five filters (in the \textit{ugriz} photometric system). An artificial neural network, ANNz2, is then trained to estimate the photometric redshift. The work in  \cite{Li2021} uses three tree-based machine learning algorithms trained on photometric measurements to extract the photometric redshift of quasars. The work in \cite{Ramachandra2022} uses a simulated dataset and trains a neural network, SYTH-Z, on photometric data to predict photometric redshift. It uses input from five bands (\textit{ugriz}), which are combined before being processed by the neural network. The work in \cite{Chunduri} uses \acp{FCN} for redshift estimation, comparing them with tree-based machine learning models.  Still, these works use tabular data rather than entire images.  \acp{CNN} are starting to be used to predict redshift directly from
images to capture additional spatial and morphological information for improved redshift estimation. In \cite{Henghes2022}, the authors use \acp{CNN} to estimate the redshift on 64 $\times$ 64 pixel images of galaxies. They use a \ac{CNN} taking in input images with 5 channels (\textit{ugriz}), but they do not use explainability to interpret the model's predictions. The architecture is formed by two convolutional layers alternated with pooling layers. Then, the output of the final layer is flattened and processed by a series of dense layers to estimate the redshift value. In \cite{schuldt_photometric_2021}, the authors use NetZ, which takes as input images with 5 channels (\textit{grizy}), but still do not use explainability techniques. The architecture is formed by two convolutional layers with max pooling and three \ac{FC} layers. In \cite{dey_photometric_2022} the authors propose PhotoZ, which exploits morphological labels during training, different from other image-only pipelines. The work in \cite{DIsanto2018}  estimates the distribution of the photometric redshift from a high-dimensional image. The architecture takes images with 15 bands as input, and no explainability methods are used. The architecture consists of three convolutional layers alternated with pooling layers. The output of the last convolutional layer is used as the input for two \ac{MDN} layers in order to retrieve the probabilistic distribution of the redshift. 
Finally, \cite{architecture} presents two \ac{CNN} architectures, A1 and A3, which take as input only images to estimate redshift, and proposes \ac{SHAP} to explain the predictions. Still, none of these works considers the case of gravitationally-lensed transients, such as gravitationally-lensed supernovae, most do not use explainability techniques to clarify the predictions or decision-making processes of their models, and none provides a quantitative evaluation of the explainability algorithms' outputs. The work in \cite{PinciroliVago2023} provides four simulated datasets, containing gravitationally-lensed supernovae, and can be used as a benchmark for redshift estimation.

\subsection{Positioning with respect to the state of the art}

While recent literature has investigated the use of deep learning for redshift estimation, this work introduces advancements that distinguish \arch from existing methodologies. First, state-of-the-art works in redshift estimation rely on custom neural networks, such as NetZ and PhotoZ, or traditional \ac{ML} regressors. In contrast, \arch is the first framework to systematically adapt generic \ac{CV} architectures (specifically EfficientNet, \ac{SwinT}, and \ac{MLPm}). By using these backbones, which are standard in computer vision but underutilized in this domain, we demonstrate that generic architectures can significantly outperform domain-specific models, such as A1 and A3, without requiring specialized, handcrafted designs. Second, this is the first study to systematically evaluate redshift estimation for gravitationally lensed transients, such as lensed supernovae. Existing studies primarily focus on galaxies or quasars, while our work validates performance across heterogeneous morphologies (e.g., Einstein rings and lensed supernovae) to prepare for surveys like \ac{LSST}, where rare transient phenomena will be the targets of cosmological studies. Finally, while \ac{XAI} tools like \ac{SHAP} have been evaluated qualitatively in previous studies, \arch proposes a quantitative evaluation of explainability results. We define formal metrics (localization accuracy and the normalized distance between the most influential pixel and the astronomical objects' centers) to validate that the model is focusing on relevant areas of the images.

\section{Datasets and methods}
\label{sec:dsmeth}

\subsection{Datasets}
\label{sec:datasets}
All the datasets consist of astronomical images that contain gravitational lenses, gravitationally-lensed supernovae, and/or galaxies.  Every image is paired with a redshift. In the simulated \ac{DGL} datasets, the \ac{GT} redshift is the one used to generate the data \cite{PinciroliVago2023, Morgan2022, pinciroli_vago_multimodal_2025}, while in the real datasets the redshift is measured. The rarity of gravitational lenses, particularly for gravitationally-lensed transients, combined with the limited availability of spectroscopic redshift data, prevents the use of large real datasets for redshift estimation on gravitational lensing data. As a result, simulated datasets are commonly employed, especially in studies involving gravitationally-lensed transients. All the datasets are divided into a training set, used to update the models' parameters, a validation set, used for early stopping and hyperparameters selection, and a test set.

\subsubsection{DeepGraviLens}

The four simulated \ac{DGL} datasets are introduced in \cite{PinciroliVago2023}, with the simulation details described more extensively in \cite{Morgan2022}, and the datasets made publicly available at \cite{zenodo}.

The datasets consist of images categorized into four classes:
\begin{itemize}
    \item No Lens: images containing only non-lensed objects, such as stars and galaxies.
    \item Lens: images of galaxy-galaxy lenses, where both the source and the lens are galaxies.
    \item \ac{LSNIa}: lenses with Type Ia supernovae as the background source. These supernovae result from the thermonuclear explosion of a white dwarf (see also \cite{Turatto2003}).
    \item \ac{LSNcc}: lenses with core-collapse supernovae as the background source, typically originating from the collapse of massive stars (see also \cite{Turatto2003}).
\end{itemize}

This work uses images from the classes Lens, \ac{LSNIa}, and \ac{LSNcc} to estimate the redshift. The class "No Lens" has been excluded, as it does not include gravitational lenses.

\begin{figure}
    \centering
    \begin{subfigure}{0.49\textwidth}
        \centering
        \includegraphics[width=\linewidth]{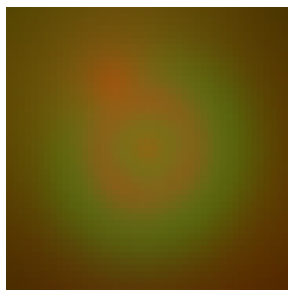}
        \caption{DES-deep}
    \end{subfigure}
    \begin{subfigure}{0.49\textwidth}
        \centering
        \includegraphics[width=\linewidth]{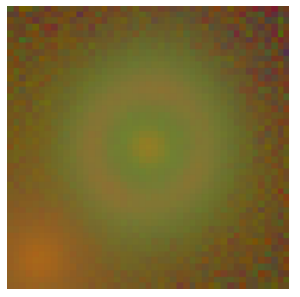}
        \caption{LSST-wide}
    \end{subfigure}
    \begin{subfigure}{0.49\textwidth}
        \centering
        \includegraphics[width=\linewidth]{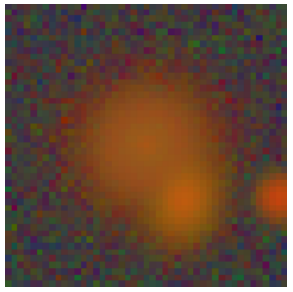}
        \caption{DESI-DOT}
    \end{subfigure}
    \begin{subfigure}{0.49\textwidth}
        \centering
        \includegraphics[width=\linewidth]{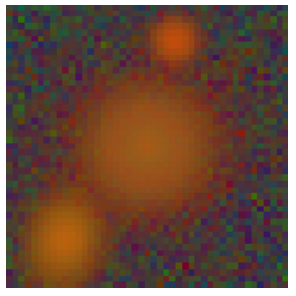}
        \caption{DES-wide}
    \end{subfigure}
    \caption{Examples of Einstein ring images in the four \ac{DGL} datasets. DES-deep, DESI-DOT and DES-wide have a side of $\approx 12$ arcsec, and LSST-wide has a side of $\approx 9$ arcsec.}
    \label{fig:example_image}
\end{figure}

\ac{DGL} comprises four simulated datasets (see also \autoref{fig:example_image}):
\begin{itemize}
    \item LSST-wide: images from \ac{LSST}, acquired with the LSSTCam camera \cite{Stalder2020}. Each pixel corresponds to 0.2 arcseconds. The exposure time is 30 seconds. The simulation reflects observational conditions estimated for the first year of the \ac{LSST} survey \cite{Marshall}.
    
    \item DESI-DOT: images from the DECam instrument \cite{Flaugher_2015}, based on real observing conditions from the \ac{DES} wide-field survey \cite{Abbott2018}. The exposure time is 60 seconds.
    
    \item DES-deep: images from DECam \cite{Flaugher_2015}, following the observing conditions of the \ac{DES} SN program \cite{Abbott2019} and has an exposure time of 200 seconds.
    
    \item DES-wide: images from DECam \cite{Flaugher_2015}, under the real observing conditions of the \ac{DES} wide-field survey \cite{Abbott2018}. The exposure time is 90 seconds.
\end{itemize}

DESI-DOT, DES-deep and DES-wide have the same scale of 0.263 arcseconds per pixel. All datasets contain 45 $\times$ 45 pixels images with 4 channels in the \textit{griz} photometric system and are divided into $70\%$ for training ($\approx10000$ samples), $15\%$ for validation ($\approx2200$ samples) and the last $15\%$ for testing ($\approx2200$ samples). \autoref{fig:example_image} shows an example image from each dataset, each featuring an Einstein ring. Since the \ac{DGL} datasets are entirely simulated, each sample is unique, thus preventing data leakage across  subsets.

\subsubsection{KiDS}

\label{sec:kids_dataset}

The \ac{KiDS} dataset contains images retrieved from the Kilo-Degree Survey \cite{Kuijken2019}, using the \ac{ESO} Science Archive Facility \cite{ESOArchive}. It comprises both low-probability and high-probability lenses. We use 4-channel images in the \textit{ugri} photometric system because of the limited availability of images with the $z$ channel in the archive. To ensure consistency between samples with and without gravitational lenses, the observations have AB magnitudes (i.e., standardized logarithmic measurements of brightness) between 20 and 25.5. The images were cropped to 45 $\times$ 45 pixels and have 0.2 arcsec/pixel, similar to the \ac{DGL} datasets. 

\begin{figure}
    \centering
    \includegraphics[width=0.5\linewidth]{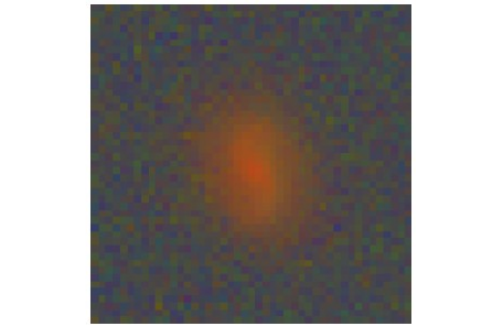}
    \caption{Example of high-probability lens in the \ac{KiDS} dataset. Each side corresponds to $\approx 9$ arcsec.}
    \label{fig:confirmed}
\end{figure}

The dataset consists of $\approx$ 4,500 candidate low-probability lenses and a subset of 264 high-probability lenses, similar to \cite{Grespan2024}. The low-probability lenses are split into training ($\approx 70\%$), validation ($\approx 15\%$), and test ($\approx 15\%$) sets. The high-probability lenses are used as an additional test set (the "lens set"). \autoref{fig:confirmed} shows an example image in the lens set. Each image is associated with a photometric redshift, estimated as presented in \cite{Baldry2014}, because of the scarcity of spectroscopic redshift measurements in the lens set. Each image is centered on a different position, thus no duplicates can be found in the dataset. Nearby objects may occasionally appear in images assigned to both the test set and the training or validation sets. However, these instances refer to distinct central objects, each associated with a different redshift. To assess potential data leakage, we considered the set of 137 couples of nearby objects, defined as those separated by angular distances compatible with an image dimension ($\approx 13$ arcsec). Then, for each couple we analyzed the correlation between the distances and the absolute differences in their redshifts. The resulting Pearson correlation coefficient is negligible ($\approx 0.06$), confirming that the presence of nearby sources does not introduce data leakage.

\subsubsection{SDSS}

\begin{table}
    \centering
    \begin{tabular}{lc}
        \toprule
        \textbf{Filters} & \textbf{Spectral range} \\
        \midrule
        u & 327-383 nm \\
        g & 406-539 nm \\
        r & 561-697 nm \\
        i & 682-838 nm \\
        \bottomrule
    \end{tabular}
    \caption{Bands of the channels used for the \ac{SDSS} dataset.}
    \label{tab:sdss_filters}
\end{table}

The \ac{SDSS} dataset comprises the data introduced in \cite{pasquet_photometric_2019}, consisting of galaxies observed in five channels corresponding to the \textit{ugriz} filter system. This dataset is used to benchmark and compare the performance of proposed architectures against the current state-of-the-art methods for redshift estimation on non-lensed galaxies. Every image is paired with the spectroscopic redshift as \ac{GT}. The dataset is divided into $\approx80\%$ train set, $\approx10\%$ validation set, and $\approx10\%$ test set, consistently with the implementation presented in \cite{architecture}, available on \url{https://github.com/1ArgoS1/PhotoZ} (as of July 2025). The images were cropped to 45$\times$ 45 $\times$ 5 pixels for consistency with the other datasets presented in this work. Each image is centered on a source of interest as presented in \cite{pasquet_photometric_2019}. The lack of data leakage is guaranteed by the uniqueness of the analyzed sources.

\begin{figure}
    \centering
    \includegraphics[width=0.5\linewidth]{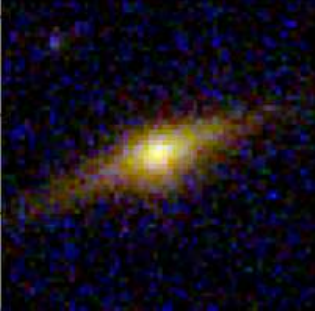}
    \caption{Example of an image taken from SDSS. Each side corresponds to $\approx 18$ arcsec.}
    \label{fig:galaxy_example}
\end{figure}

\subsection{Tasks, targets and outputs}
The main task of this work is redshift estimation from astronomical observations and is formulated as a regression problem from multi-band 45 $\times$ 45 pixels images. Training targets are derived from either simulations or redshift measurements. The second task is explainability, which shows where the best models focus their attention. For each image and predicted redshift, we use \ac{SHAP} heatmaps to highlight the most influential pixels. 
    We perform a quantitative evaluation of the \ac{SHAP} heatmaps by comparing the most influential regions against the position of the gravitational lenses. First, we define the ground truth bounding boxes. These are generated automatically by computing the maximum intensity across image channels and selecting the pixels with the top 10\% intensity values. The resulting bounding boxes are then manually validated. Then, we analyze the \ac{SHAP} heatmaps. We identify the most influential pixel in each image as the global maximum of each \ac{SHAP} heatmap. We then compute two metrics: the distance between this peak value and the lens center, and the localization accuracy (i.e., the percentage of times the peak \ac{SHAP} points fall within the bounding boxes). Overall, high localization accuracy and low mean and median values indicate that the models focus on the most relevant parts of the images, while low localization accuracy and high mean and median values either indicate that the model does not focus on the correct parts of the images or that the object of interest spreads across the entire image (as in the case of DES-deep).

\begin{figure}
    \centering
    \begin{subfigure}[b]{0.49\textwidth}
        \includegraphics[width=\linewidth]{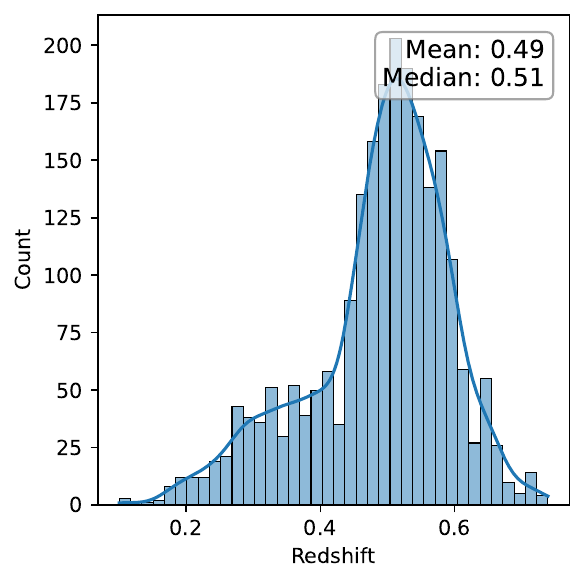}
        \caption{\ac{DGL}}
    \end{subfigure}
    \hfill
    \begin{subfigure}[b]{0.49\textwidth}
        \includegraphics[width=\linewidth]{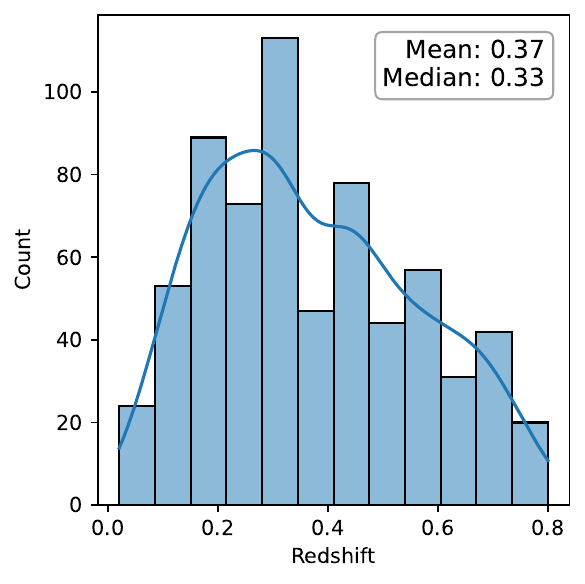}
        \caption{\ac{KiDS} (test)}
    \end{subfigure}
    \vskip\baselineskip
    \begin{subfigure}[b]{0.49\textwidth}
        \includegraphics[width=\linewidth]{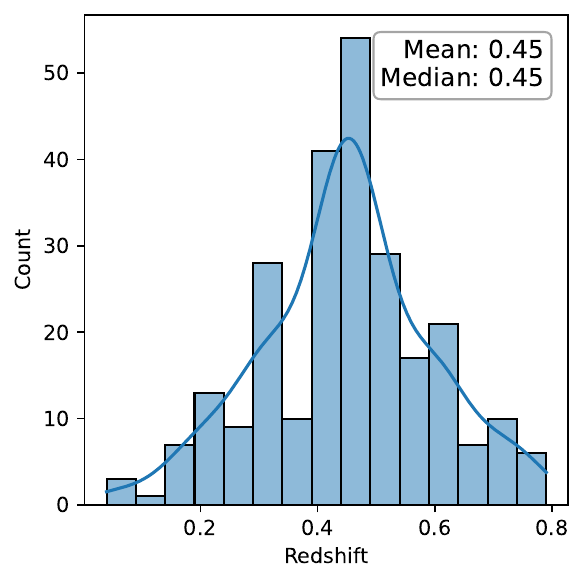}
        \caption{\ac{KiDS} (lens)}
    \end{subfigure}
    \hfill
    \begin{subfigure}[b]{0.49\textwidth}
        \includegraphics[width=\linewidth]{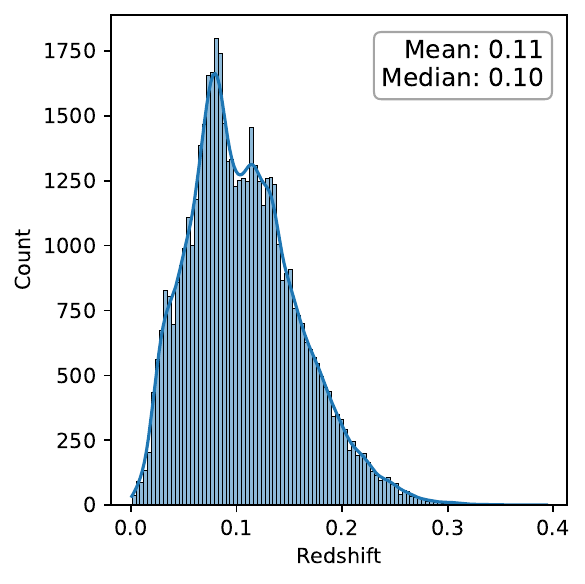}
        \caption{\ac{SDSS}}
    \end{subfigure}
    \caption{\ac{GT} redshift distributions for \ac{DGL}, \ac{KiDS} and \ac{SDSS}.}
    \label{fig:redshift_distributions}
\end{figure}

\autoref{fig:redshift_distributions} compares the \ac{GT} redshift distributions for \ac{DGL} (a - the four datasets have the same distribution), \ac{KiDS} (b and c), and \ac{SDSS} (d). The four distributions have different mean values (ranging from 0.11 to 0.49), median values (ranging from 0.10 to 0.51), and skewnesses (negative for \ac{DGL}, positive for \ac{KiDS} test set and \ac{SDSS} and no skewness for \ac{KiDS} lens set). These differences allow the evaluation of the models' robustness across diverse scenarios.

\subsection{Loss functions and metrics}
The \ac{MSE} is defined as:
\begin{equation}
     MSE=\frac{1}{N}\sum{(y_i-\hat{y_i})^2}
\end{equation}
where $N$ is the number of samples, $y_i$ is the label of the sample $i$ (i.e., the \ac{GT} redshift) and $\hat{y_i}$ is the value predicted by the model for sample $i$. \ac{MSE} penalizes larger errors heavily due to the squaring of residuals, discouraging large deviations from \ac{GT} values.

\ac{MAE} is defined as:
\begin{equation}
    MAE=\frac{1}{N}\sum{|y_i-\hat{y_i}|}
\end{equation}

Different from \ac{MSE}, \ac{MAE} uses the same units as the redshift.

$R^2$ is defined as:
\begin{equation}
    R^2=1-\frac{\sum{(y_i-\hat{y_i})^2}}{\sum{(y_i-\overline{y_i})^2}}
    \label{eq:rsquare}
\end{equation}

where $\overline{y_i}$ is the average value of the redshift in the dataset. $R^2$ is a standard metrics for regression problems (see e.g. \cite{hoekendijk_counting_2021, kochukrishnan_comprehensive_2024}). The closer the $R^2$ is to 1, the better the estimator is because it indicates a low estimation error compared to the variance.

The \ac{NMAD} is a robust estimator of the variance, defined as:
\begin{equation}
    \text{NMAD} = 1.4826 \cdot \text{median}\left(\left| \frac{\hat{y_i}-y_i}{1+y_i} \right|\right)
\end{equation}

Unlike standard deviation, \ac{NMAD} is less sensitive to catastrophic outliers, making it a preferred metric for photometric redshift performance.

$\sigma_{68}$ is defined based on the percentiles of the normalized residuals $\Delta z_i = \frac{\hat{y_i}-y_i}{1+y_i}$:
\begin{equation}
    \sigma_{68} = \frac{P_{84}(\Delta z) - P_{16}(\Delta z)}{2}
\end{equation}
where $P_{84}$ and $P_{16}$ are the 84th and 16th percentiles of the distribution of $\Delta z$, respectively. Similar to \ac{NMAD}, $\sigma_{68}$ provides a robust measure of the dispersion that corresponds to one standard deviation for a normal distribution, ignoring the influence of extreme tails.

The prediction bias is defined as the median of the normalized residuals:
\begin{equation}
    \text{Bias} = \text{median}\left(\frac{\hat{y_i}-y_i}{1+y_i}\right)
\end{equation}
where $y_i$ is the \ac{GT} redshift and $\hat{y_i}$ is the predicted value. This metric quantifies the systematic offset of the model predictions. A positive bias indicates a tendency to overestimate the redshift, while a negative bias implies underestimation.

The outlier fraction ($\eta$) quantifies the percentage of samples with "catastrophic" prediction errors, defined as:
\begin{equation}
    \eta = \frac{1}{N} \sum_{i=1}^{N} \mathbb{I}\left(\left| \frac{\hat{y_i}-y_i}{1+y_i} \right| > \theta \right)
\end{equation}
where $\mathbb{I}$ is the indicator function which equals 1 if the condition is met and 0 otherwise, and $\theta$ is the outlier threshold (set to $0.05$ in this work). This metric assesses the reliability of the model by tracking the frequency of outliers.

Section \ref{sec:training} reports which losses and early stopping metrics are used for each dataset.

\subsubsection{Uncertainty estimation}
\label{sec:uncertainty_estimation}
Uncertainty estimation allows the identification of significant improvements in metrics for different experiments (see e.g.,  \cite{PinciroliVago2023}). In this work, $1\sigma$ uncertainties are computed on all the evaluation metrics and are estimated on the test set using bootstrapping, a 4-step statistical technique:

\begin{enumerate}
\item Starting from a test set containing $D$ samples (each consisting of a \ac{GT} label and its associated prediction), create a bootstrap sample $b_i$ by drawing $d$ samples randomly with replacement from the original set.
\item Calculate the desired evaluation metric (e.g., \ac{MAE}), denoted as $m_i$, on the resampled set $b_i$.
\item Repeat steps 1 and 2 a total of $k$ times to generate a collection of $k$ metric values: $\{m_1, m_2, ..., m_k\}$.
\item Estimate the mean and uncertainty (here reported as $\pm 1\sigma$) from the resulting distribution of metric values.
\end{enumerate}

In this work, the number of data points in the resampled test set is $d = D$ (see also \cite{PinciroliVago2023}), and the number of bootstrapping iterations is $k = 1000$.

\subsection{Pipeline}

\begin{figure}
    \centering
    \includegraphics[width=\linewidth]{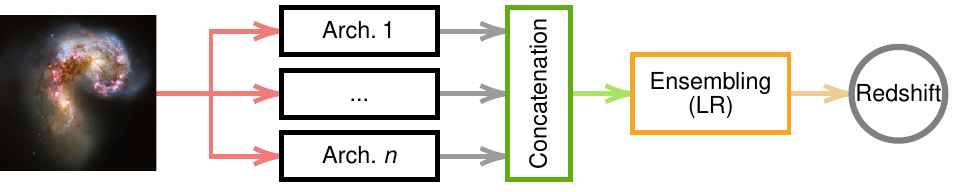}
    \caption{The \arch pipeline}
    \label{fig:pipeline}
\end{figure}

This work proposes \arch, an \ac{ML}-based pipeline, to estimate the redshift from astronomical images. \autoref{fig:pipeline} illustrates the pipeline. First, \ac{ML} and \ac{DL} architectures (\textit{Arch. 1} to \textit{Arch. n}) are trained. Then, the estimations of the best architectures are ensembled using a \ac{LRE} on the best architectures' output latent vectors. The \ac{ML} and \ac{DL} architectures belong to different families:
\begin{itemize}
    \item Feature extraction: this family comprises \ac{HOG} \cite{HOG} + \ac{SVR}\footnote{As implemented in \url{https://scikit-learn.org/stable/modules/generated/sklearn.svm.SVR.html}.}
    \item \ac{CNN}: this family comprises \ac{ResNet} \cite{resnet}, EfficientNet \cite{efficientnet}, A1 \cite{architecture}, A3 \cite{architecture}, NetZ \cite{schuldt_photometric_2021} and PhotoZ \cite{dey_photometric_2022}
    \item \ac{MLP}: this family comprises \ac{MLPm}, a non-convolutional \ac{NN} based on \acf{MLP} \cite{mlp}
    \item Transformers: this family comprises \ac{SwinT} \cite{swin}
\end{itemize}

The presented architectures, with the exception of A1, A3, NetZ and PhotoZ, were initially designed for classification. In this work, they are adapted for the regression task by replacing the classification head with a regression head (similar to \cite{regressor}) formed by a single \ac{FC} layer.

\subsubsection{HOG + SVR}
\ac{HOG} is a \ac{CV} algorithm designed to capture the edges and shapes of objects in an image. It begins by computing horizontal and vertical gradients using predefined filter kernels, from which the gradient magnitude and orientation are derived for each pixel. The image is then divided into non-overlapping cells (5 $\times$ 5 pixels in this work), and each cell is represented by a histogram of gradient orientations weighted by magnitude. These cells are grouped into blocks. The final image representation is formed by concatenating and normalizing the features of all blocks. To reduce dimensionality and mitigate overfitting, \ac{PCA} is applied to the resulting feature vector \cite{Jolliffe2016}. Additional features are obtained by computing the histograms of colours from the original images. The resulting features are given as input to \ac{SVR}, which is trained to estimate redshift. \autoref{tab:hyper} presents the relevant hyperparameters and their values. The other hyperparameters have their default values. "Color features" indicates whether the histograms of colours are retained, "Number of orientations (\ac{HOG})" indicates the number of discrete values of the gradient orientations, "Explained variance (\ac{PCA})" indicates the minimum required explained variance for \ac{PCA} and "$\epsilon$ (\ac{SVR})" defines a margin of tolerance within which prediction errors are not penalized by the loss function. Overall, the experiments consider 24 hyperparameter combinations.

\begin{table}
    \centering
\caption{Hyperparameters of the \ac{HOG} + \ac{SVR} architecture.}
    \begin{tabular}{ll}
        \toprule
        \textbf{Hyperparameter} & \textbf{Values} \\
        \midrule
        Color features           & yes, no          \\
        Number of orientations (\ac{HOG})  & 8, 16            \\
        Explained variance (\ac{PCA})      & 0.6, 0.8, 0.9    \\
        $\epsilon$ (\ac{SVR})              & 0.1, 0.01        \\
        \bottomrule
    \end{tabular}

    \label{tab:hyper}
\end{table}

\subsubsection{ResNet}

\ac{ResNet} \cite{resnet} is a \ac{DL} architecture introduced to improve stability and convergence during training. This architecture has introduced residual blocks, which are the key to stabilization and prevent the problem of vanishing gradients \cite{Shafiq2022}. In residual blocks, the input is directly added to the output of the block through a residual connection. The network\footnote{As implemented in \url{https://huggingface.co/docs/transformers/en/model_doc/resnet} (as of July 2025)} is formed by four stages of residual blocks. The output of the fourth block is given as input to the \ac{FC} layers, which form the regression head. \autoref{tab:resnet} presents the non-default hyperparameters and their values, for a total of 54 hyperparameter combinations. The embedding size indicates the number of dimensions in which the images are represented in the feature space. A small value can make the network unable to learn sufficient information, while a high value could lead to overfitting. The $L_2$ regularization coefficient controls regularization by adding a penalty to the loss function proportional to the square of the model's parameters. This reduces the risk of overfitting by discouraging large weights. The \ac{LR} determines how much model weights are updated during training: high values may prevent convergence to the global optimum, while overly low values can lead to premature convergence in local minima. The depth hyperparameter defines the number of layers for each stage.

\begin{table}
\centering
\caption{Hyperparameters for the \ac{ResNet} architecture.}
\begin{tabular}{ll}
\toprule
\textbf{Hyperparameter} & \textbf{Values} \\
\midrule
Embedding size & 16, 64, 256 \\
$L_2$ regularization coefficient        & 0.01, 0.001, 0.0001 \\
\ac{LR}        & 0.01, 0.001, 0.0001 \\
Depth     & {[}2,2,2,2{]}, {[}3,4,6,3{]} \\
\bottomrule
\end{tabular}
\label{tab:resnet}
\end{table}

\subsubsection{EfficientNet}
EfficientNet \cite{efficientnet}\footnote{As implemented in \url{https://huggingface.co/docs/transformers/model_doc/efficientnet} (as of July 2025)} is a \ac{CNN}-based architecture that uses compound scaling to uniformly scale its dimensions (width, depth, and resolution) with fixed coefficients, rather than scaling each of them separately. Width refers to the number of channels, depth refers to the number of layers, and resolution refers to the size of the input data. In this work, the resolution is fixed and the width and depth scaling parameters are set as in the official Tensorflow implementation\footnote{Available at \url{https://github.com/tensorflow/tpu/tree/master/models/official/efficientnet} (as of July 2025)}. This work uses the architectures from B1 to B3 (detailed in \autoref{tab:efficientnet_type}) as more complex architectures are more prone to overfitting. \autoref{tab:hyperparam_efficient} lists the hyperparameters and their corresponding values, showing a total of 54 combinations.

\begin{table}
\centering
\caption{Details of the EfficientNet architectures B1 to B3 \cite{efficientnet}}
\begin{tabular}{ccc}
\toprule
\textbf{Architecture} & \textbf{Width Coefficient} & \textbf{Depth Coefficient} \\
\midrule
B1 & 1.0               & 1.1               \\
B2 & 1.1               & 1.2               \\
B3 & 1.2               & 1.4               \\
\bottomrule
\end{tabular}
\label{tab:efficientnet_type}
\end{table}

\begin{table}
\centering
\caption{Hyperparameters for EfficientNet.}
\begin{tabular}{ll}
\toprule
\textbf{Hyperparameter} & \textbf{Values} \\
\midrule
Dropout & 0.3, 0.5 \\
$L_2$ regularization coefficient      & 0.01, 0.001, 0.0001 \\
\ac{LR}      & 0.01, 0.001, 0.0001 \\
Architecture type    & B1, B2, B3 \\
\bottomrule
\end{tabular}
\label{tab:hyperparam_efficient}
\end{table}

\subsubsection{A1}
A1 is a \ac{CNN} designed for regression tasks presented in \cite{architecture}.  It comprises two convolutional layers followed by a PReLU activation and average pooling. The resulting feature map is flattened and passed to three \ac{FC} layers, each followed by a PReLU activation.

\subsubsection{A3}
A3 \cite{architecture} is a \ac{CNN} architecture designed for regression tasks, incorporating Inception blocks to capture multi-scale features. Its structure includes an initial convolutional layer followed by average pooling, two Inception blocks, another average pooling layer, a third Inception block, a flattening layer, and finally three \ac{FC} layers.

\subsubsection{PhotoZ}
PhotoZ \cite{pasquet_photometric_2019} is a deep \ac{CNN} architecture that uses Inception blocks. Its structure includes an initial convolutional layer followed by a sequence of five Inception blocks designed to capture multi-scale spatial features, a flattening layer, and finally multiple \ac{FC} layers to produce the redshift prediction.

\subsubsection{NetZ}
NetZ \cite{schuldt_photometric_2021} is a \ac{CNN} architecture designed to estimate photometric redshifts directly from multi-band galaxy images. Its structure comprises a series of convolutional layers for feature extraction, followed by max-pooling layers to reduce spatial dimensions, and \ac{FC} layers to output the redshift prediction.

\subsubsection{MLPm}
\ac{MLPm} \cite{mlp} is a \ac{NN} that relies on \acp{MLP} to process spatial and channel-wise information instead of using convolutions or attention-based mechanisms. In this architecture, the input image is first divided into non-overlapping patches, which are then converted into vectors and projected into a higher-dimensional space using a linear layer; this process forms the patch embeddings. The main component of the network is the Mixer block, which alternates between two \acp{MLP}: one for channel mixing and one for spatial mixing. The channel mixing \ac{MLP} mixes information across feature channels independently for each spatial location. The spatial (or token) mixing \ac{MLP}, on the other hand, mixes information across feature channels independently for each feature channel. \autoref{tab:mlp_type} summarizes the variants of \ac{MLPm} (detailed in \cite{mlp}), and \autoref{tab:mlp_hyperparameters} presents the hyperparameters and their values, for a total of 108 combinations.

\begin{table}
\centering
\caption{Variants of the \ac{MLPm} architectures \cite{mlp}.}
\label{tab:mlp_type}
\begin{tabular}{ccc}
\toprule
\textbf{Architecture} & \textbf{Hidden size} & \textbf{Number of layers} \\
\midrule
S & 512 & 8 \\
B & 768 & 12 \\
L & 1024 & 24 \\
\bottomrule
\end{tabular}
\end{table}

\begin{table}
    \centering
    \caption{Hyperparameters considered for the tuning of the \ac{MLPm} architecture.}
    \label{tab:mlp_hyperparameters}
    \begin{tabular}{ll}
        \toprule
        \textbf{Hyperparameter} & \textbf{Values} \\
        \midrule
        \ac{LR}         & 0.01, 0.001, 0.0001 \\
        $L_2$ regularization coefficient         & 0.01, 0.001, 0.0001 \\
        Patch size & 5, 9 \\
        Architecture variant       & S, B, L \\
        Dropout    & 0.0, 0.2 \\
        \bottomrule
    \end{tabular}
\end{table}

\subsubsection{Swin transformer}
\ac{SwinT} is an architecture based on the transformer block. It employs self-attention to capture multi-scale representations of images. Rather than applying attention globally, it partitions the image into patches and groups them into windows. As a result, the model can capture both local and global contextual information. This work uses a window size of 2 patches and a patch size of 5 pixels. The transformer's 4-layer hierarchical structure is used to look for general information, while the smaller window size allows for the focus on fine-grained local features. The work in \cite{swin} shows that the model generalizes better than the original \ac{ViT} on datasets with fewer samples. For instance, in \cite{Han2024}, \ac{SwinT} has been used for healthcare data. \autoref{tab:swin_type} shows the architecture variants, and \autoref{tab:swin_hyper} shows the hyperparameters' values, for a total of 162 combinations. In this work, we consider only the architectures with a small number of parameters to prevent overfitting.

\begin{table}
\centering
\caption{Architecture variants for \ac{SwinT} \cite{swin}.}
\label{tab:swin_type}
\begin{tabular}{cccc}
\toprule
\textbf{Architecture} & \textbf{Embedding size} & \textbf{Number of heads} & \textbf{Depth} \\
\midrule
T & 96  & [3, 6, 12, 24] & [2, 2, 6, 2] \\
S & 96  & [3, 6, 12, 24] & [2, 2, 18, 2] \\
B & 128 & [4, 8, 16, 32] & [2, 2, 18, 2] \\
\bottomrule
\end{tabular}
\end{table}

\begin{table}
\centering
\caption{Hyperparameters considered for the tuning of the Swin transformer architecture.}
\label{tab:swin_hyper}
\begin{tabular}{ll}
\toprule
\textbf{Hyperparameter} & \textbf{Values} \\
\midrule
Dropout        & 0.0, 0.2 \\
Encoder stride & 3, 5, 9 \\
$L_2$ regularization coefficient             & 0.01, 0.001, 0.0001 \\
\ac{LR}             & 0.01, 0.001, 0.0001 \\
Architecture variant           & T, S, B \\
\bottomrule
\end{tabular}
\end{table}

\subsubsection{Ensembling}

Ensembling takes as input the predictions from different base architectures and outputs a new prediction. All possible combinations of 2, 3, and 4 different architectures were tested to identify the best combination for each dataset. The best base models were selected based on their validation set \ac{MAE} (i.e., for each architecture, we selected the best hyperparameters as the ones associated with the lowest validation \ac{MAE}).

We use \ac{LRE} to combine the predictions. This algorithm determines the final prediction by finding a linear combination of the base architectures' predictions \cite{breiman_stacked_1996}, as follows:

\begin{equation}
    y=\beta_0+ \sum_{i=1}^N \beta_ix_i
\end{equation}

where $y$ is the output of the \ac{LRE} model, $\beta_0$ is a constant value, $x_i$ are the outputs of the selected input architectures, $\beta_i$ are their coefficients and $N$ is the number of selected input architectures. The function is optimized by minimizing the \ac{MSE}.

\subsubsection{Training}
\label{sec:training}

The training process of \arch is divided into two stages. First, all the architectures are trained separately, using the same inputs. Then \ac{LRE}, is trained on the outputs of 2 to 4 of the input architectures. Identical data splits are used in both stages to avoid double-dipping.

\begin{table}[ht]
\centering
\caption{Training and stopping criteria per dataset}
\label{tab:summary_losses}
\begin{tabular}{ccccc}
\toprule
\multirow{2}{*}{\textbf{Dataset}} & \multirow{2}{*}{\textbf{Loss function}} & \multirow{2}{*}{\textbf{Max epochs}} & \multicolumn{2}{c}{\textbf{Early stopping}} \\ \cmidrule(lr){4-5}
 &  &  & \textbf{Patience} & \textbf{Metrics} \\ \midrule
\textbf{\ac{DGL}} & \ac{MSE} & 500 & 100 & $R^2$ \\
\textbf{\ac{KiDS}} & \ac{MSE} & 500 & 100 & $R^2$ \\
\textbf{\ac{SDSS}} & \ac{MAE} & 100 & 20 & \ac{MAE} \\ \bottomrule
\end{tabular}
\end{table}

Table \ref{tab:summary_losses} reports the loss function, maximum number of epochs and early stopping parameters for all the datasets. All the input architectures for \ac{DGL} and \ac{KiDS} are trained for 500 epochs with early stopping on the $R^2$ (which, different from \ac{MAE} and \ac{MSE}, is scale-invariant and is a direct indicator of the model's explanatory power) and a patience of 100 epochs, similar to \cite{PinciroliVago2023}. For each dataset, the best model is the one with the best \ac{MAE} on the validation set. For the \ac{SDSS} dataset, all architectures are trained for 100 epochs (consistently with \cite{architecture}) with a patience of 20. \ac{MAE} is used for both the loss function and early stopping in \arch \ac{DL} architectures, as it outperforms the combination of \ac{MSE} for the loss and $R^2$ for early stopping (see also \autoref{sec:sdss_results}). The best hyperparameters for each network are chosen based on the validation set and tested on a lockbox test set. The best architecture-hyperparameter combinations are also chosen on the validation set and tested on the same lockbox test set.

\subsection{Explainability with SHAP}

\ac{SHAP} is a game-theoretic approach used to explain \ac{ML} models predictions. When applied to images, it identifies the regions of the images that contribute most significantly to the model's predictions.

The work in \cite{Dardouillet2023} provides a summary of the algorithm for image analysis. \ac{SHAP} relies on a masker to hide portions of the images. In this work, the masker blurs patches of the image using a 5 $\times$ 5 pixels kernel. The pre-trained regression model used for estimating the redshift is then applied to the masked images, and the \ac{SHAP} value is computed for each masked region $i$ as follows:

    \begin{equation}
        \phi_i = \sum_{S\subseteq N\setminus\{i\}}\frac{|S|!\,(|N|-|S|-1)!}{|N|!}\,\left[f(S\cup\{i\})-f(S)\right]
    \end{equation}

    where $\phi_i$ is the \ac{SHAP} value for region $i$, $S$ is a subset of regions that does not include $i$, $N$ is the complete set of features (or regions), $f(S)$ is the result of the model with the region $i$ blurred, $f(S\cup\{i\})$ is the result of the model considering $S$ plus the region $i$ not blurred and $\frac{|S|!\,(|N|-|S|-1)!}{|N|!}$ is the weight that normalizes the contribution of $i$ based on the size of $S$.
  
  Finally, the \ac{SHAP} values ($\phi_i$) of each region are aggregated to provide a heatmap for each image. In this work, the most influential regions are those with the highest absolute values of the \ac{SHAP} coefficients ($|\phi_i|$).

\section{Evaluation}
\label{sec:results}

This section reports the quantitative and qualitative evaluation of \arch on the datasets introduced in \autoref{sec:datasets}. For each result, uncertainty is reported as presented in \autoref{sec:uncertainty_estimation}. In this work, \arch architectures (\ac{ResNet}, \ac{MLPm}, \ac{SwinT}, EfficientNet and ensembling) are compared with baseline architectures (\ac{HOG} + \ac{SVR}, A1, A3, NetZ and PhotoZ). The \ac{DL} baseline architectures are those presented in \cite{architecture,schuldt_photometric_2021,pasquet_photometric_2019}, and \ac{HOG} + \ac{SVR} is a \ac{ML} baseline.

\subsection{DeepGraviLens}

\begin{table}[]
\centering
\caption{\ac{MAE} on the test set for each dataset in \ac{DGL}. All numerical results are rounded to the third decimal place for consistency. The best results, and those falling within their confidence intervals, are highlighted in bold.}
\label{tab:dgl_mae}
\begin{tabular}{@{}ccccc@{}}
\toprule
\textbf{Network} & \textbf{DES-deep} & \textbf{DES-wide} & \textbf{DESI-DOT} & \textbf{LSST-wide} \\ \midrule
\textbf{HOG + SVR} & 0.067 $\pm$ 0.001 & 0.045 $\pm$ 0.001 & 0.052 $\pm$ 0.001 & 0.048 $\pm$ 0.001 \\
\textbf{A1} & 0.064 $\pm$ 0.001 & 0.024 $\pm$ 0.000 & 0.029 $\pm$ 0.001 & 0.027 $\pm$ 0.001 \\
\textbf{A3} & 0.061 $\pm$ 0.001 & 0.033 $\pm$ 0.001 & 0.037 $\pm$ 0.001 & 0.036 $\pm$ 0.001 \\
\textbf{NetZ} & 0.064 $\pm$ 0.001 & 0.023 $\pm$ 0.001 & 0.027 $\pm$ 0.001 & 0.025 $\pm$ 0.001 \\
\textbf{PhotoZ} & 0.061 $\pm$ 0.001 & 0.013 $\pm$ 0.000 & 0.020 $\pm$ 0.001 & 0.016 $\pm$ 0.000 \\
\textbf{EfficientNet} & \textbf{0.034 $\pm$ 0.001} & 0.013 $\pm$ 0.000 & 0.016 $\pm$ 0.000 & 0.014 $\pm$ 0.000 \\
\textbf{MLPm} & 0.053 $\pm$ 0.001 & 0.010 $\pm$ 0.000 & 0.015 $\pm$ 0.000 & 0.013 $\pm$ 0.000 \\
\textbf{ResNet} & 0.069 $\pm$ 0.001 & 0.010 $\pm$ 0.000 & 0.017 $\pm$ 0.000 & 0.016 $\pm$ 0.000 \\
\textbf{SwinT} & 0.045 $\pm$ 0.001 & 0.010 $\pm$ 0.000 & 0.014 $\pm$ 0.000 & 0.012 $\pm$ 0.000 \\ \midrule
\textbf{Ensemble} & \textbf{0.034 $\pm$ 0.001} & \textbf{0.009 $\pm$ 0.000} & \textbf{0.012 $\pm$ 0.000} & \textbf{0.011 $\pm$ 0.000} \\ \midrule
\textbf{Improvement {[}\%{]}} & 45 & 36 & 38 & 32 \\ \bottomrule
\end{tabular}
\end{table}

\begin{table}[]
\centering
\caption{\ac{NMAD} on the test set for each dataset in \ac{DGL}. All numerical results are rounded to the third decimal place for consistency. The best results, and those falling within their confidence intervals, are highlighted in bold.}
\label{tab:dgl_nmad}
\begin{tabular}{@{}ccccc@{}}
\toprule
 & \textbf{DES-deep} & \textbf{DES-wide} & \textbf{DESI-DOT} & \textbf{LSST-wide} \\ \midrule
\textbf{HOG + SVR} & 0.051 $\pm$ 0.002 & 0.037 $\pm$ 0.001 & 0.040 $\pm$ 0.001 & 0.036 $\pm$ 0.001 \\
\textbf{A1} & 0.051 $\pm$ 0.001 & 0.017 $\pm$ 0.000 & 0.022 $\pm$ 0.001 & 0.020 $\pm$ 0.001 \\
\textbf{A3} & 0.048 $\pm$ 0.001 & 0.027 $\pm$ 0.001 & 0.029 $\pm$ 0.001 & 0.027 $\pm$ 0.001 \\
\textbf{NetZ} & 0.051 $\pm$ 0.001 & 0.016 $\pm$ 0.000 & 0.021 $\pm$ 0.001 & 0.018 $\pm$ 0.000 \\
\textbf{PhotoZ} & 0.048 $\pm$ 0.002 & 0.007 $\pm$ 0.000 & 0.011 $\pm$ 0.000 & 0.009 $\pm$ 0.000 \\
\textbf{EfficientNet} & \textbf{0.021 $\pm$ 0.001} & 0.010 $\pm$ 0.000 & 0.009 $\pm$ 0.000 & 0.009 $\pm$ 0.000 \\
\textbf{MLPm} & 0.039 $\pm$ 0.001 & 0.004 $\pm$ 0.000 & 0.006 $\pm$ 0.000 & \textbf{0.005 $\pm$ 0.000} \\
\textbf{ResNet} & 0.054 $\pm$ 0.001 & 0.005 $\pm$ 0.000 & 0.009 $\pm$ 0.000 & 0.009 $\pm$ 0.000 \\
\textbf{SwinT} & 0.032 $\pm$ 0.001 & 0.005 $\pm$ 0.000 & 0.007 $\pm$ 0.000 & 0.008 $\pm$ 0.000 \\ \midrule
\textbf{Ensemble} & 0.023 $\pm$ 0.001 & \textbf{0.003 $\pm$ 0.000} & \textbf{0.005 $\pm$ 0.000} & \textbf{0.005 $\pm$ 0.000} \\ \midrule
\textbf{Improvement {[}\%{]}} & 55 & 51 & 52 & 46 \\ \bottomrule
\end{tabular}
\end{table}

\begin{table}[]
\centering
\caption{$\sigma_{68}$ on the test set for each dataset in \ac{DGL}. All numerical results are rounded to the third decimal place for consistency. The best results, and those falling within their confidence intervals, are highlighted in bold.}
\label{tab:dgl_sigma}
\begin{tabular}{@{}ccccc@{}}
\toprule
 & \textbf{DES-deep} & \textbf{DES-wide} & \textbf{DESI-DOT} & \textbf{LSST-wide} \\ \midrule
\textbf{HOG + SVR} & 0.054 $\pm$ 0.001 & 0.036 $\pm$ 0.001 & 0.041 $\pm$ 0.001 & 0.038 $\pm$ 0.001 \\
\textbf{A1} & 0.052 $\pm$ 0.001 & 0.018 $\pm$ 0.000 & 0.022 $\pm$ 0.001 & 0.021 $\pm$ 0.001 \\
\textbf{A3} & 0.049 $\pm$ 0.001 & 0.026 $\pm$ 0.000 & 0.029 $\pm$ 0.001 & 0.028 $\pm$ 0.001 \\
\textbf{NetZ} & 0.051 $\pm$ 0.001 & 0.017 $\pm$ 0.000 & 0.021 $\pm$ 0.000 & 0.019 $\pm$ 0.000 \\
\textbf{PhotoZ} & 0.048 $\pm$ 0.001 & 0.008 $\pm$ 0.000 & 0.012 $\pm$ 0.000 & 0.010 $\pm$ 0.000 \\
\textbf{EfficientNet} & \textbf{0.024 $\pm$ 0.001} & 0.008 $\pm$ 0.000 & 0.010 $\pm$ 0.000 & 0.009 $\pm$ 0.000 \\
\textbf{MLPm} & 0.040 $\pm$ 0.001 & 0.005 $\pm$ 0.000 & \textbf{0.007 $\pm$ 0.000} & \textbf{0.006 $\pm$ 0.000} \\
\textbf{ResNet} & 0.054 $\pm$ 0.001 & 0.006 $\pm$ 0.000 & 0.011 $\pm$ 0.000 & 0.010 $\pm$ 0.000 \\
\textbf{SwinT} & 0.035 $\pm$ 0.001 & 0.005 $\pm$ 0.000 & 0.008 $\pm$ 0.000 & 0.008 $\pm$ 0.000 \\ \midrule
\textbf{Ensemble} & 0.026 $\pm$ 0.001 & \textbf{0.004 $\pm$ 0.000} & \textbf{0.007 $\pm$ 0.000} & \textbf{0.006 $\pm$ 0.000} \\ \midrule
\textbf{Improvement {[}\%{]}} & 50 & 49 & 45 & 43 \\ \bottomrule
\end{tabular}
\end{table}

\begin{table}[]
\centering
\caption{Bias on the test set for each dataset in \ac{DGL}. All numerical results are rounded to the third decimal place for consistency. The best results, and those falling within their confidence intervals, are highlighted in bold.}
\label{tab:dgl_bias}
\begin{tabular}{@{}ccccc@{}}
\toprule
 & \textbf{DES-deep} & \textbf{DES-wide} & \textbf{DESI-DOT} & \textbf{LSST-wide} \\ \midrule
\textbf{HOG + SVR} & -0.002 $\pm$ 0.001 & \textbf{0.000 $\pm$ 0.001} & \textbf{0.000 $\pm$ 0.001} & \textbf{0.000 $\pm$ 0.001} \\
\textbf{A1} & -0.002 $\pm$ 0.001 & \textbf{-0.000 $\pm$ 0.000} & -0.003 $\pm$ 0.001 & -0.001 $\pm$ 0.000 \\
\textbf{A3} & -0.004 $\pm$ 0.001 & -0.003 $\pm$ 0.001 & -0.002 $\pm$ 0.001 & -0.002 $\pm$ 0.001 \\
\textbf{NetZ} & -0.009 $\pm$ 0.001 & -0.001 $\pm$ 0.000 & -0.002 $\pm$ 0.001 & -0.003 $\pm$ 0.000 \\
\textbf{PhotoZ} & -0.009 $\pm$ 0.001 & 0.001 $\pm$ 0.000 & \textbf{0.000 $\pm$ 0.000} & \textbf{0.000 $\pm$ 0.000} \\
\textbf{EfficientNet} & \textbf{0.001 $\pm$ 0.000} & 0.001 $\pm$ 0.000 & 0.001 $\pm$ 0.000 & \textbf{0.000 $\pm$ 0.000} \\
\textbf{MLPm} & \textbf{0.001 $\pm$ 0.001} & -0.001 $\pm$ 0.000 & -0.002 $\pm$ 0.000 & \textbf{0.000 $\pm$ 0.000} \\
\textbf{ResNet} & -0.011 $\pm$ 0.001 & 0.001 $\pm$ 0.000 & 0.001 $\pm$ 0.000 & -0.005 $\pm$ 0.000 \\
\textbf{SwinT} & \textbf{-0.001 $\pm$ 0.001} & \textbf{0.000 $\pm$ 0.000} & \textbf{0.000 $\pm$ 0.000} & 0.001 $\pm$ 0.000 \\ \midrule
\textbf{Ensemble} & \textbf{-0.000 $\pm$ 0.001} & \textbf{-0.000 $\pm$ 0.000} & \textbf{0.000 $\pm$ 0.000} & \textbf{-0.000 $\pm$ 0.000} \\ \bottomrule
\end{tabular}
\end{table}

\begin{table}[]
\centering
\caption{Outlier rate on the test set for each dataset in \ac{DGL}. All numerical results are rounded to the third decimal place for consistency. The best results, and those falling within their confidence intervals, are highlighted in bold.}
\label{tab:dgl_outlier}
\begin{tabular}{@{}ccccc@{}}
\toprule
 & \textbf{DES-deep} & \textbf{DES-wide} & \textbf{DESI-DOT} & \textbf{LSST-wide} \\ \midrule
\textbf{HOG + SVR} & 0.354 $\pm$ 0.010 & 0.182 $\pm$ 0.008 & 0.232 $\pm$ 0.009 & 0.203 $\pm$ 0.009 \\
\textbf{A1} & 0.335 $\pm$ 0.010 & 0.040 $\pm$ 0.004 & 0.060 $\pm$ 0.005 & 0.056 $\pm$ 0.005 \\
\textbf{A3} & 0.307 $\pm$ 0.010 & 0.096 $\pm$ 0.006 & 0.109 $\pm$ 0.006 & 0.102 $\pm$ 0.007 \\
\textbf{NetZ} & 0.341 $\pm$ 0.010 & 0.036 $\pm$ 0.004 & 0.045 $\pm$ 0.004 & 0.044 $\pm$ 0.004 \\
\textbf{PhotoZ} & 0.319 $\pm$ 0.010 & 0.017 $\pm$ 0.003 & 0.041 $\pm$ 0.004 & 0.032 $\pm$ 0.004 \\
\textbf{EfficientNet} & \textbf{0.114 $\pm$ 0.007} & 0.018 $\pm$ 0.003 & 0.025 $\pm$ 0.003 & 0.021 $\pm$ 0.003 \\
\textbf{MLPm} & 0.253 $\pm$ 0.009 & 0.020 $\pm$ 0.003 & 0.042 $\pm$ 0.004 & 0.036 $\pm$ 0.004 \\
\textbf{ResNet} & 0.368 $\pm$ 0.010 & 0.013 $\pm$ 0.002 & 0.031 $\pm$ 0.004 & 0.027 $\pm$ 0.003 \\
\textbf{SwinT} & 0.194 $\pm$ 0.008 & 0.017 $\pm$ 0.003 & 0.025 $\pm$ 0.003 & 0.021 $\pm$ 0.003 \\ \midrule
\textbf{Ensemble} & \textbf{0.111 $\pm$ 0.007} & \textbf{0.009 $\pm$ 0.002} & \textbf{0.019 $\pm$ 0.003} & \textbf{0.012 $\pm$ 0.002} \\ \midrule
\textbf{Improvement {[}\%{]}} & 64 & 48 & 54 & 61 \\ \bottomrule
\end{tabular}
\end{table}

\autoref{tab:dgl_mae} summarizes the results in terms of \ac{MAE} and shows that the choice of the architecture is determinant. \ac{HOG} + \ac{SVR}, a simple architecture, is able to exploit some shape information to estimate redshift, but is significantly worse than all the \ac{DL} architectures. State-of-the-art \ac{DL} algorithms and ensembling yields substantial improvements (32\% to 45\%) over baseline \ac{DL} architectures (A1, A3, NetZ and PhotoZ) and \ac{HOG} + \ac{SVR}. Overall, non-ensembled algorithms already outperform the baselines, while ensembling improves performance further or matches that of the best individual architecture (e.g., DES-deep). The worst architecture among those presented in this work is \ac{ResNet}. Overall, while ensembling is the best method on all the datasets, results change significantly based on the datasets' characteristics, as also shown in  \cite{PinciroliVago2023}. DES-deep is the most challenging dataset, leading to higher \ac{MAE}. Moreover, the architectures presented in this work achieve a larger improvement in \ac{MAE} over the best baseline architecture ($\approx 45\%$) compared with other datasets ($\approx 35\%$), suggesting that for more noisy datasets the features extracted by deep architectures are more relevant than in higher-quality datasets.

The better results of the \arch \ac{DL} algorithms presented in this work are also observed when comparing the $R^2$ scores. In DES-deep, the \arch \ac{DL} algorithms have, on average, $R^2 \approx 0.60$, compared with $R^2 \approx 0.31$ of \ac{HOG} + \ac{SVR} and $R^2 \approx 0.38$ for the five baseline architectures. In DES-wide, $R^2 \approx 0.97$ for the \arch \ac{DL} architectures, compared to an average of $\approx 0.85$ for the baseline architectures. In DESI-DOT, the \arch architectures have $R^2 \approx 0.94$, compared to an average of $R^2 \approx 0.80$ for the baseline algorithms. In LSST, $R^2 \approx 0.95$ for the \arch architectures, compared to an average baseline of $R^2 \approx 0.82$.

Similar conclusions can be drawn from \autoref{tab:dgl_nmad} and \autoref{tab:dgl_sigma}, which report the \ac{NMAD} and $\sigma_{68}$ dispersion metrics, respectively. They confirm the superiority of the proposed \ac{DL} architectures. \ac{MLPm} and ensembling consistently achieve the lowest dispersion across the DES-wide, DESI-DOT, and LSST-wide datasets. Notably, on the challenging DES-deep dataset, EfficientNet has the lowest scatter ($\sigma_{68} \approx 0.024$ and \ac{NMAD} $\approx 0.021$), outperforming ensembling slightly. The improvements in scatter reduction are substantial (46\% to 55\% for \ac{NMAD} and 43\% to 50\% for $\sigma_{68}$) compared to the best baseline methods. A paired bootstrap test (16th–84th percentile confidence interval) on \ac{NMAD} confirms a statistically significant improvement of our best architecture over the baselines.

\autoref{tab:dgl_bias} shows that the bias is negligible for the majority of the architectures and datasets. The \ac{DL} models presented in this work mitigate systematic offsets, achieving bias values of $\approx 0$ across all datasets. While the baseline \ac{DL} models (e.g., NetZ, PhotoZ) and \ac{HOG} + \ac{SVR} occasionally lead to small biases (up to -0.009 in DES-deep), the proposed architectures are on average less biased.

Finally, \autoref{tab:dgl_outlier} highlights the significant reduction in the outlier rate. On the higher-quality datasets (DES-wide, DESI-DOT, LSST-wide), the proposed architectures remove most of the catastrophic failures, achieving an outlier rate $< 2\%$. Even on the more challenging DES-deep dataset, the outlier rate is reduced by $\approx 64\%$ compared to the baselines, achieving $\approx 11\%$ for ensembling and EfficientNet. These improvements further demonstrate the robustness of the proposed models.

\subsubsection{Explainability}

\begin{figure}
    \centering
    \includegraphics[width=\linewidth]{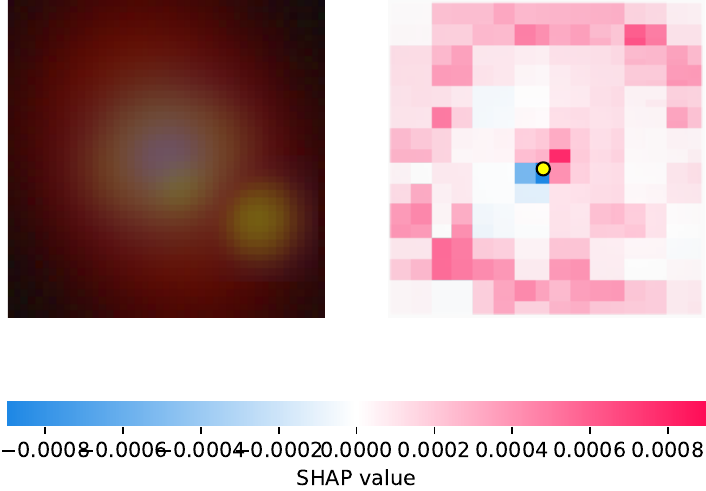}
    \caption{Example of \ac{SHAP} heatmap on the DES-deep dataset. 
    The yellow dot indicates the position of the most influential pixel according to \ac{SHAP}. Each side corresponds to $\approx 12$ arcsec.}
    \label{fig:dgl_shap_desdeep}
\end{figure}

\begin{figure}
    \centering
    \includegraphics[width=\linewidth]{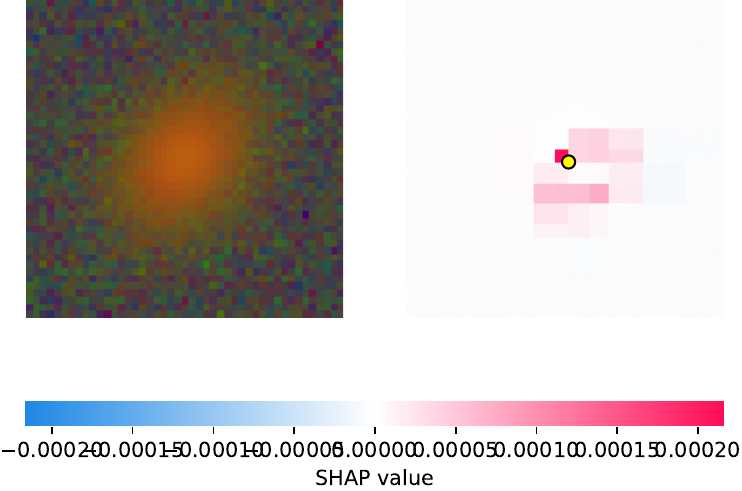}
    \caption{Example of \ac{SHAP} heatmap on the DES-wide dataset. 
    The yellow dot indicates the position of the most influential pixel according to \ac{SHAP}. Each side corresponds to $\approx 12$ arcsec.}
    \label{fig:dgl_shap_deswide}
\end{figure}

\begin{figure}
    \centering
    \includegraphics[width=\linewidth]{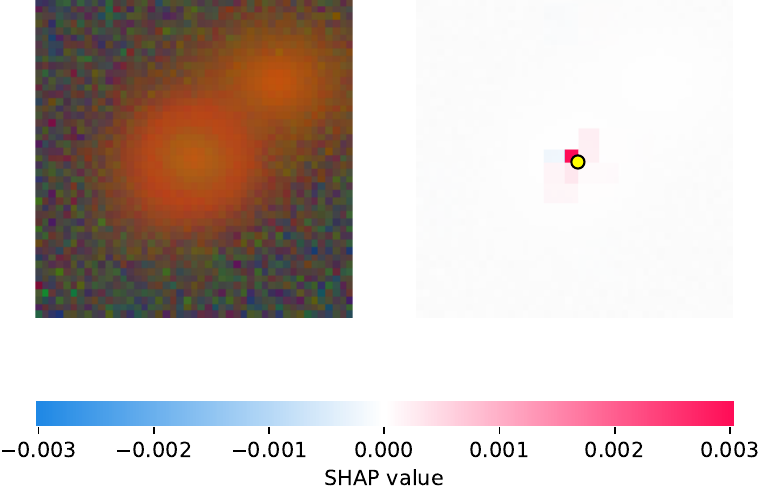}
    \caption{Example of \ac{SHAP} heatmap on the DESI-DOT dataset. 
    The yellow dot indicates the position of the most influential pixel according to \ac{SHAP}. Each side corresponds to $\approx 12$ arcsec.}
    \label{fig:dgl_shap_desidot}
\end{figure}

\begin{figure}
    \centering
    \includegraphics[width=\linewidth]{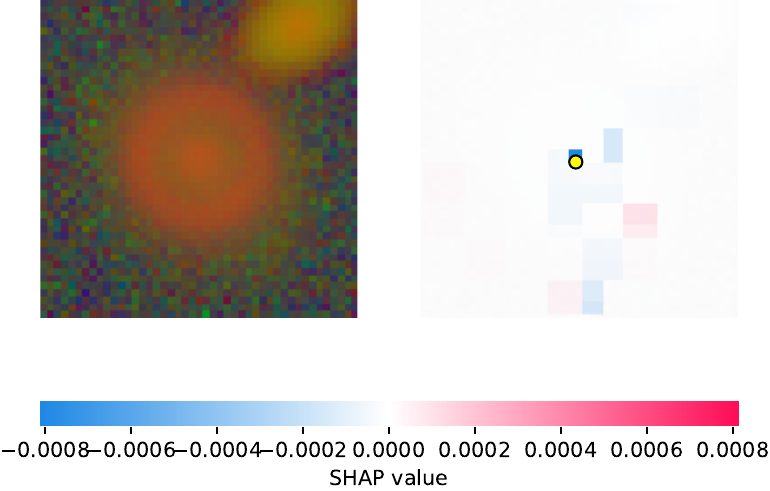}
    \caption{Example of \ac{SHAP} heatmap on the LSST-wide dataset. 
    The yellow dot indicates the position of the most influential pixel according to \ac{SHAP}. Each side corresponds to $\approx 9$ arcsec.}
    \label{fig:dgl_shap_lsst}
\end{figure}

Figures \ref{fig:dgl_shap_desdeep}, \ref{fig:dgl_shap_deswide}, \ref{fig:dgl_shap_desidot}, and \ref{fig:dgl_shap_lsst} show representative examples of \ac{SHAP} heatmaps for lensing systems in \ac{DGL}, obtained from the best non-ensembled models. In each image, the yellow dot indicates the position of the region associated with the highest absolute \ac{SHAP} value.
DES-deep is the most challenging dataset due to the noisy background, so \ac{SHAP} assigns high importance to multiple regions, reflecting uncertainty in identifying the features determining redshift. In contrast, DES-wide shows a more localised \ac{SHAP} heatmap, concentrated on the lensing system, where the model is focusing. The DESI-DOT example exhibits a sharp peak, implying that the model relies almost exclusively on the lensing object and not on the Einstein ring or the background. Lastly, in LSST-wide, the influential regions are more spread, indicating that the model uses both the lens and the background objects to estimate redshift.

\begin{figure}
    \centering
    \begin{subfigure}[b]{0.49\textwidth}
        \includegraphics[width=\linewidth]{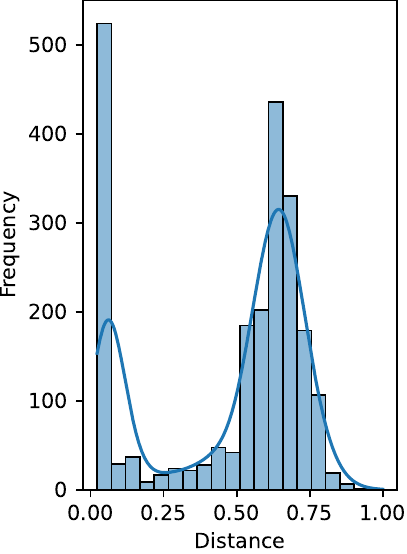}
        \caption{DES-deep}
    \end{subfigure}
    \hfill
    \begin{subfigure}[b]{0.49\textwidth}
        \includegraphics[width=\linewidth]{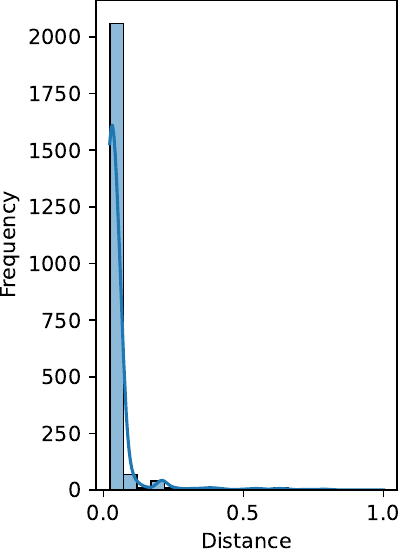}
        \caption{DES-wide}
    \end{subfigure}
    \hfill
    \begin{subfigure}[b]{0.49\textwidth}
        \includegraphics[width=\linewidth]{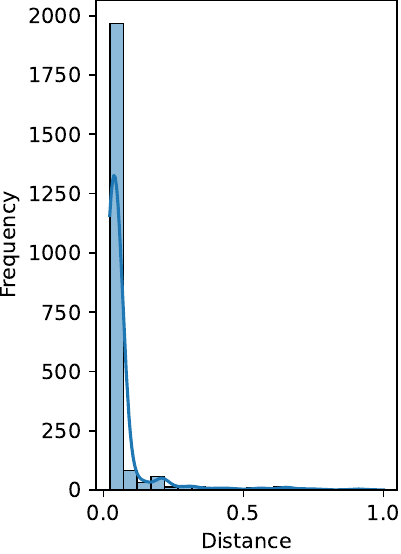}
        \caption{DESI-DOT}
    \end{subfigure}
    \hfill
    \begin{subfigure}[b]{0.49\textwidth}
        \includegraphics[width=\linewidth]{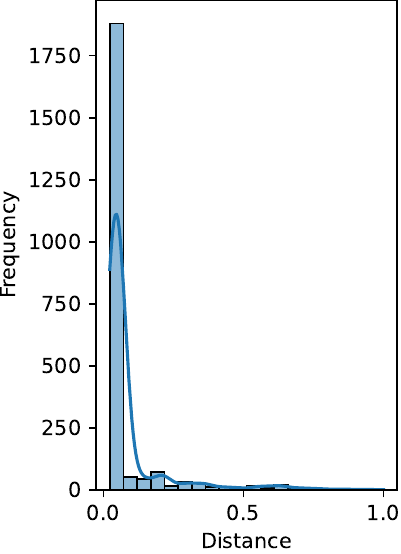}
        \caption{LSST-wide}
    \end{subfigure}
    \caption{Distribution of the distances between the most influential pixels and the centres of the images for the four \ac{DGL} dataset. The distances range between 0 and 1, where 1 indicates the maximum possible distance in a single image.}
    \label{fig:dgl_shap_distributions}
\end{figure}

To assess the ability of the model to focus on the most relevant areas, we calculate the distance between the most influential region and the centre of the image (where the object of interest is also centred). Each distance is normalized between 0 and 1. \autoref{fig:dgl_shap_distributions} shows the distribution of the distances for the four datasets. The mean distance for DES-deep is 0.474 (median 0.598), indicating that the model's focus diverges significantly from the lensing region and that the lens spreads over the image. In contrast, the distances are significantly lower for LSST-wide (mean 0.090, median 0.050), DESI-DOT (mean 0.068, median 0.050), and DES-wide (mean 0.054, median 0.022), confirming that the model typically concentrates its attention near the lens centre. This trend suggests that higher image quality leads to more interpretable behaviour, improving trustworthiness in the redshift estimation models. Table \ref{tab:dgl_loc} reports the localization accuracy for the \ac{DGL} datasets. As observed in distances distribution, the most challenging dataset is DES-deep, with a localization accuracy of $\approx 38\%$, while the localization accuracy is $> 95\%$ for the other \ac{DGL} datasets.

The nature of \ac{KiDS} targets introduces a challenge regarding the reliability of \ac{GT} labels (i.e., photometric redshifts derived from a broader set of spectral channels than the four bands available to the \arch models). This discrepancy creates an information gap, as the targets contain information that is absent in the input images. As a consequence, the \ac{ML} algorithms estimate redshift but also approximate the outputs of a higher-dimensional estimator using limited photometric data. This reliance on estimated targets introduces label noise, but it also imposes a pessimistic performance ceiling compared to the higher-dimensional case, which contains more information. The proposed architectures achieve low bias and good performance across different metrics despite the information deficit, which demonstrates their capability to extract robust features even from limited observational data.

\begin{table}[h]
    \centering
    \caption{Localization accuracy for the \ac{DGL} datasets.}
    \label{tab:dgl_loc}
    \begin{tabular}{lcc}
        \toprule
        \textbf{Dataset} & \textbf{Bounding Box Size (\%)} & \textbf{Localization Accuracy (\%)} \\
        \midrule
        DES-deep  & 17.00 & 38.08 \\
        DES-wide  & 18.00 & 99.07 \\
        DESI-DOT  & 18.19 & 97.92 \\
        LSST-wide & 19.01 & 95.11 \\
        \bottomrule
    \end{tabular}
\end{table}

\subsection{KiDS}

\autoref{tab:real_mae} summarizes the results of the proposed architectures in terms of \ac{MAE}. As in the case of \ac{DGL}, deep architectures outperform \ac{HOG} + \ac{SVR}, indicating that shape-based information is not sufficient for accurate redshift estimation. In general, state-of-the-art architectures outperform simpler baseline architectures such as A1, A3, PhotoZ and NetZ, and ensembling is, for most of the metrics, as effective or more effective than individual architectures. Its effectiveness is observed consistently in the two subsets (where "test" refers to the lower-probability candidate lenses and "lens" to the high-probability lenses, as presented in \autoref{sec:kids_dataset}). Both \ac{SwinT} and \ac{MLPm}, which are not based on \acp{CNN}, outperform the other architectures. Still, \ac{MAE} is higher than on \ac{DGL} data, suggesting that the noise in real data makes estimations more difficult. Furthermore, the best \ac{MAE} ($> 0.06$) is greater than the error introduced by the use of photometric redshift in place of spectroscopic redshift ($\approx 0.01$) presented in \cite{wright_kids-legacy_2025}. For this reason, the reported improvements reflect an improvement in the model's ability to estimate redshift, rather than overfitting to label noise. Overall, the resulting \ac{MAE} can be regarded as an upper bound of the error, given by the error introduced by \ac{GT} photometric redshift and the estimation error introduced by the network.

As in the case of the simulated datasets, the better results of the \arch \ac{DL} algorithms are also observed when comparing the $R^2$ score. For the test dataset, $R^2 \approx 0.72$, compared with $R^2 \approx 0.61$ for the baseline architectures. For high-probability lenses, $R^2 \approx 0.62$ for the new architectures, compared with $R^2 \approx 0.45$ for the baseline ones. In this case, the result is more informative than \ac{MAE} alone, as the distribution of the redshifts is different between the two subsets, and high-probability lenses tend to have a higher redshift in this dataset.

\autoref{tab:kids_nmad} and \autoref{tab:kids_sigma} present the results for the dispersion metrics \ac{NMAD} and $\sigma_{68}$, respectively. The proposed \ac{DL} architectures consistently achieve lower scatter compared to the baselines on both the "lens" and "test" subsets. Ensembling, \ac{SwinT} and \ac{MLPm} perform particularly well on average. The improvement in scatter reduction is significant, reaching up to 27\% for \ac{NMAD} and 17\% for $\sigma_{68}$ compared to the best baseline models. While the absolute values are higher than those in the simulated \ac{DGL} datasets, this is expected given the higher noise levels in real data. A paired bootstrap test (16th–84th percentile confidence interval) on \ac{NMAD} confirms a statistically significant improvement of our best architecture over the baselines. 

\autoref{tab:kids_bias} shows that the proposed architectures mitigate bias. Ensembling, EfficientNet, \ac{MLPm}, \ac{ResNet}, and \ac{SwinT} all achieve low bias values. In contrast, some baseline models, such as A1 and PhotoZ, lead to small but non-negligible biases, especially on the lens set (up to 0.017). This confirms the robustness of the proposed methods in producing unbiased estimates even on real observational data.

Finally, \autoref{tab:kids_outlier} highlights the reduction in outliers. The outlier rate is significantly improved by the proposed architectures, with Ensembling achieving the lowest rate on both the lens set ($\approx 0.296$) and the test set ($\approx 0.289$). This represents an improvement of up to $\approx 16\%$ compared to the best baselines. Although the outlier rates are higher than in the simulated data, the relative improvement demonstrates the capability of these models to handle real observations better than existing methods.

\begin{table}[]
\centering
\caption{\ac{MAE} of the models trained on the \ac{KiDS} dataset. The results are taken on the test set and lens set for the models with the combination of hyperparameters with the best performance on the validation set.  All numerical results are rounded to the third decimal place for consistency. The best results, and those falling within their confidence intervals, are highlighted in bold.}
\label{tab:real_mae}
\begin{tabular}{@{}ccc@{}}
\toprule
\textbf{Network} & \textbf{\ac{KiDS} (lens)} & \textbf{\ac{KiDS} (test)} \\ \midrule
\textbf{HOG + SVR} & 0.108 $\pm$ 0.006 & 0.114 $\pm$ 0.004 \\
\textbf{A1} & 0.072 $\pm$ 0.004 & 0.072 $\pm$ 0.003 \\
\textbf{A3} & 0.073 $\pm$ 0.004 & 0.077 $\pm$ 0.003 \\
\textbf{NetZ} & 0.069 $\pm$ 0.004 & 0.069 $\pm$ 0.003 \\
\textbf{PhotoZ} & 0.070 $\pm$ 0.004 & 0.071 $\pm$ 0.003 \\
\textbf{EfficientNet} & 0.077 $\pm$ 0.004 & 0.068 $\pm$ 0.003 \\
\textbf{MLPm} & \textbf{0.064 $\pm$ 0.004} & \textbf{0.066 $\pm$ 0.003} \\
\textbf{ResNet} & 0.066 $\pm$ 0.005 & 0.071 $\pm$ 0.003 \\
\textbf{SwinT} & \textbf{0.064 $\pm$ 0.004} & 0.067 $\pm$ 0.003 \\ \midrule
\textbf{Ensemble} & \textbf{0.060 $\pm$ 0.004} & \textbf{0.063 $\pm$ 0.003} \\ \midrule
\textbf{Improvement {[}\%{]}} & 14 & 10 \\ \bottomrule
\end{tabular}
\end{table}

\begin{table}[]
\centering
\caption{\ac{NMAD} of the models trained on the \ac{KiDS} dataset. The results are taken on the test set and lens set for the models with the combination of hyperparameters with the best performance on the validation set.  All numerical results are rounded to the third decimal place for consistency. The best results, and those falling within their confidence intervals, are highlighted in bold.}
\label{tab:kids_nmad}
\begin{tabular}{@{}ccc@{}}
\toprule
 & \textbf{KiDS (lens)} & \textbf{KiDS (test)} \\ \midrule
\textbf{HOG + SVR} & 0.081 $\pm$ 0.006 & 0.088 $\pm$ 0.005 \\
\textbf{A1} & 0.051 $\pm$ 0.004 & 0.049 $\pm$ 0.003 \\
\textbf{A3} & 0.051 $\pm$ 0.005 & 0.054 $\pm$ 0.004 \\
\textbf{NetZ} & 0.049 $\pm$ 0.004 & 0.046 $\pm$ 0.003 \\
\textbf{PhotoZ} & 0.052 $\pm$ 0.004 & 0.049 $\pm$ 0.002 \\
\textbf{EfficientNet} & 0.052 $\pm$ 0.004 & 0.046 $\pm$ 0.003 \\
\textbf{MLPm} & 0.041 $\pm$ 0.003 & 0.044 $\pm$ 0.002 \\
\textbf{ResNet} & \textbf{0.040 $\pm$ 0.004} & 0.050 $\pm$ 0.002 \\
\textbf{SwinT} & 0.042 $\pm$ 0.004 & 0.043 $\pm$ 0.002 \\ \midrule
\textbf{Ensemble} & \textbf{0.036 $\pm$ 0.004} & \textbf{0.039 $\pm$ 0.002} \\ \midrule
\textbf{Improvement {[}\%{]}} & 27 & 16 \\ \bottomrule
\end{tabular}
\end{table}

\begin{table}[]
\centering
\caption{$\sigma_{68}$ of the models trained on the \ac{KiDS} dataset. The results are taken on the test set and lens set for the models with the combination of hyperparameters with the best performance on the validation set.  All numerical results are rounded to the third decimal place for consistency. The best results, and those falling within their confidence intervals, are highlighted in bold.}
\label{tab:kids_sigma}
\begin{tabular}{@{}ccc@{}}
\toprule
 & \textbf{KiDS (lens)} & \textbf{KiDS (test)} \\ \midrule
\textbf{HOG + SVR} & 0.078 $\pm$ 0.007 & 0.096 $\pm$ 0.004 \\
\textbf{A1} & 0.054 $\pm$ 0.004 & 0.057 $\pm$ 0.003 \\
\textbf{A3} & 0.058 $\pm$ 0.005 & 0.061 $\pm$ 0.003 \\
\textbf{NetZ} & 0.051 $\pm$ 0.004 & 0.055 $\pm$ 0.003 \\
\textbf{PhotoZ} & 0.054 $\pm$ 0.005 & 0.054 $\pm$ 0.004 \\
\textbf{EfficientNet} & \textbf{0.050 $\pm$ 0.005} & 0.051 $\pm$ 0.003 \\
\textbf{MLPm} & \textbf{0.049 $\pm$ 0.003} & 0.049 $\pm$ 0.003 \\
\textbf{ResNet} & \textbf{0.050 $\pm$ 0.004} & 0.054 $\pm$ 0.003 \\
\textbf{SwinT} & \textbf{0.049 $\pm$ 0.005} & 0.048 $\pm$ 0.003 \\ \midrule
\textbf{Ensemble} & \textbf{0.045 $\pm$ 0.005} & \textbf{0.045 $\pm$ 0.003} \\ \midrule
\textbf{Improvement {[}\%{]}} & 13 & 17 \\  \bottomrule
\end{tabular}
\end{table}

\begin{table}[]
\centering
\caption{Bias of the models trained on the \ac{KiDS} dataset. The results are taken on the test set and lens set for the models with the combination of hyperparameters with the best performance on the validation set.  All numerical results are rounded to the third decimal place for consistency. The best results, and those falling within their confidence intervals, are highlighted in bold.}
\label{tab:kids_bias}
\begin{tabular}{@{}ccc@{}}
\toprule
 & \textbf{KiDS (lens)} & \textbf{KiDS (test)} \\ \midrule
\textbf{HOG + SVR} & -0.010 $\pm$ 0.007 & \textbf{0.000 $\pm$ 0.004} \\
\textbf{A1} & 0.017 $\pm$ 0.004 & \textbf{0.002 $\pm$ 0.002} \\
\textbf{A3} & 0.014 $\pm$ 0.003 & 0.004 $\pm$ 0.003 \\
\textbf{NetZ} & -0.010 $\pm$ 0.003 & 0.004 $\pm$ 0.002 \\
\textbf{PhotoZ} & 0.015 $\pm$ 0.003 & 0.004 $\pm$ 0.002 \\
\textbf{EfficientNet} & -0.024 $\pm$ 0.002 & -0.007 $\pm$ 0.002 \\
\textbf{MLPm} & \textbf{0.002 $\pm$ 0.004} & -0.004 $\pm$ 0.002 \\
\textbf{ResNet} & \textbf{-0.001 $\pm$ 0.003} & 0.006 $\pm$ 0.002 \\
\textbf{SwinT} & -0.011 $\pm$ 0.003 & \textbf{-0.001 $\pm$ 0.002} \\ \midrule
\textbf{Ensemble} & \textbf{-0.004 $\pm$ 0.003} & \textbf{-0.000 $\pm$ 0.002} \\ \bottomrule
\end{tabular}
\end{table}

\begin{table}[]
\centering
\caption{Outlier rate of the models trained on the \ac{KiDS} dataset. The results are taken on the test set and lens set for the models with the combination of hyperparameters with the best performance on the validation set.  All numerical results are rounded to the third decimal place for consistency. The best results, and those falling within their confidence intervals, are highlighted in bold.}
\label{tab:kids_outlier}
\begin{tabular}{@{}ccc@{}}
\toprule
 & \textbf{KiDS (lens)} & \textbf{KiDS (test)} \\ \midrule
\textbf{HOG + SVR} & 0.539 $\pm$ 0.031 & 0.556 $\pm$ 0.019 \\
\textbf{A1} & 0.379 $\pm$ 0.031 & 0.359 $\pm$ 0.019 \\
\textbf{A3} & 0.387 $\pm$ 0.031 & 0.395 $\pm$ 0.019 \\
\textbf{NetZ} & 0.333 $\pm$ 0.029 & 0.349 $\pm$ 0.019 \\
\textbf{PhotoZ} & 0.355 $\pm$ 0.030 & 0.343 $\pm$ 0.019 \\
\textbf{EfficientNet} & 0.399 $\pm$ 0.031 & 0.331 $\pm$ 0.019 \\
\textbf{MLPm} & \textbf{0.323 $\pm$ 0.029} & 0.319 $\pm$ 0.018 \\
\textbf{ResNet} & 0.330 $\pm$ 0.030 & 0.331 $\pm$ 0.018 \\
\textbf{SwinT} & 0.332 $\pm$ 0.029 & 0.310 $\pm$ 0.018 \\ \midrule
\textbf{Ensemble} & \textbf{0.296 $\pm$ 0.028} & \textbf{0.289 $\pm$ 0.018} \\ \midrule
\textbf{Improvement {[}\%{]}} & 11 & 16 \\ \bottomrule
\end{tabular}
\end{table}

\subsubsection{Explainability}

\begin{figure}
    \centering
    \includegraphics[width=\linewidth]{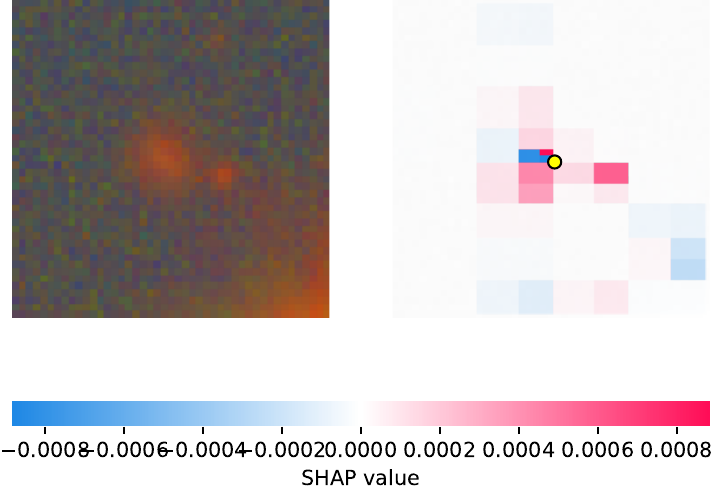}
    \caption{Example of a \ac{SHAP} heatmap on the \ac{KiDS} dataset (test). 
    The yellow dot indicates the position of the most influential pixel according to \ac{SHAP}. Each side corresponds to $\approx 9$ arcsec.}
    \label{fig:kids_shap_test_414}
\end{figure}

\begin{figure}
    \centering
    \includegraphics[width=\linewidth]{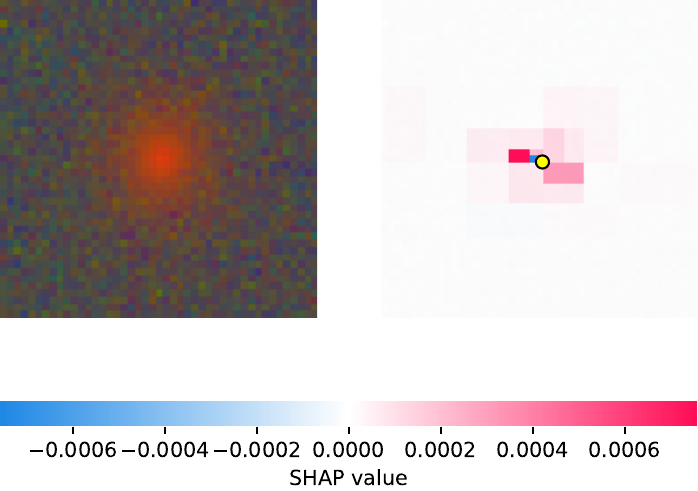}
    \caption{Example of a \ac{SHAP} heatmap on the \ac{KiDS} dataset (test). 
    The yellow dot indicates the position of the most influential pixel according to \ac{SHAP}. Each side corresponds to $\approx 9$ arcsec.}
    \label{fig:kids_shap_test_380}
\end{figure}

\begin{figure}
    \centering
    \includegraphics[width=\linewidth]{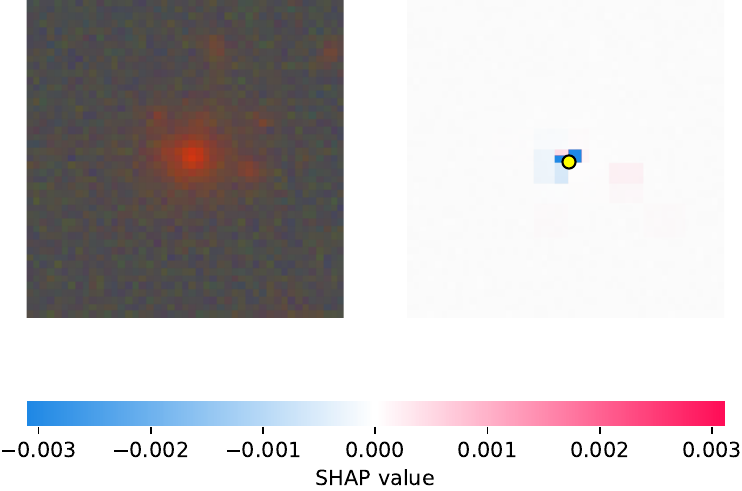}
    \caption{Example of a \ac{SHAP} heatmap on the \ac{KiDS} dataset (lens). 
    The yellow dot indicates the position of the most influential pixel according to \ac{SHAP}. Each side corresponds to $\approx 9$ arcsec.}
    \label{fig:kids_shap_lens_102}
\end{figure}

\begin{figure}
    \centering
    \includegraphics[width=\linewidth]{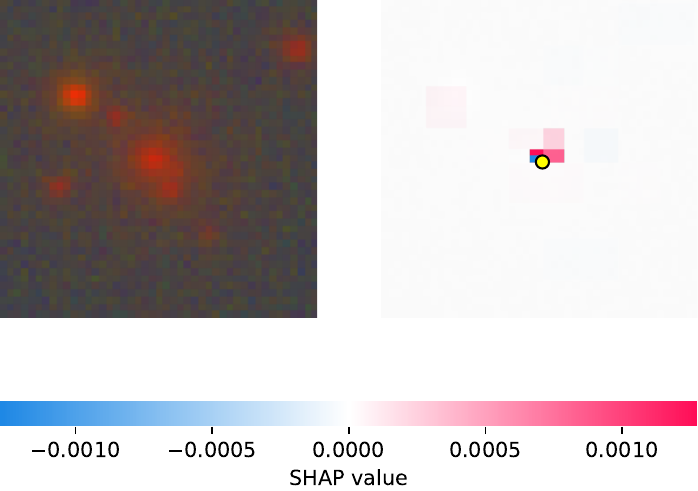}
    \caption{Example of a \ac{SHAP} heatmap on the \ac{KiDS} dataset (lens). 
    The yellow dot indicates the position of the most influential pixel according to \ac{SHAP}. Each side corresponds to $\approx 9$ arcsec.}
    \label{fig:kids_shap_lens_20}
\end{figure}

Figures \ref{fig:kids_shap_test_414}, \ref{fig:kids_shap_test_380}, \ref{fig:kids_shap_lens_102}, and  \ref{fig:kids_shap_lens_20} show representative examples of \ac{SHAP} heatmaps for images in \ac{KiDS}, obtained from the best non-ensembled models. In each image, the yellow dot indicates the pixel with the highest absolute \ac{SHAP} value (i.e., the most influential region for the model's estimate).
Low-probability lens images (Figures \ref{fig:kids_shap_test_414}, \ref{fig:kids_shap_test_380}) are associated with \ac{SHAP} heatmaps that exhibit low-magnitude values dispersed around regions near the centre, suggesting more uncertainty in the most relevant region for redshift estimation. In contrast, Figures \ref{fig:kids_shap_lens_102} and  \ref{fig:kids_shap_lens_20} correspond to high-probability lens candidates, and \ac{SHAP} heatmaps are more concentrated near the image centers.

\begin{figure}
    \centering
    \begin{subfigure}[b]{0.49\textwidth}
        \includegraphics[width=\linewidth]{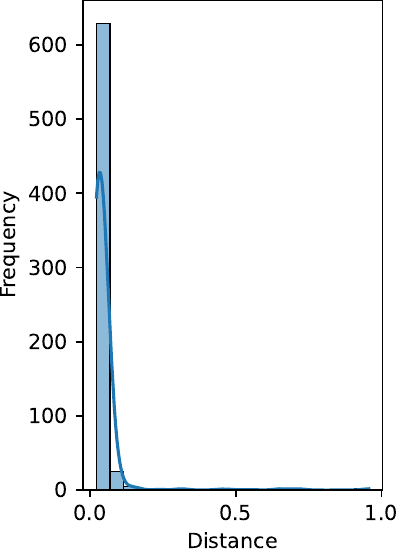}
        \caption{\ac{KiDS} (test)}
    \end{subfigure}
    \hfill
    \begin{subfigure}[b]{0.49\textwidth}
        \includegraphics[width=\linewidth]{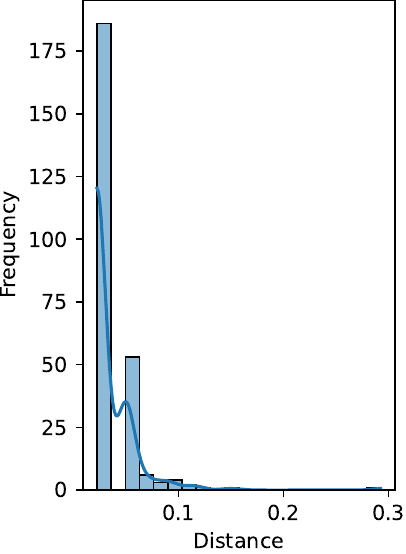}
        \caption{\ac{KiDS} (lens)}
    \end{subfigure}
    \caption{Distribution of the distances between the most influential pixels and the centres of the images for the \ac{KiDS} dataset. The distances range between 0 and 1, where 1 indicates the maximum possible distance in a single image.}
    \label{fig:kids_shap_distributions}
\end{figure}

To quantify the relevance of the most influential pixels, \autoref{fig:kids_shap_distributions} shows the distribution of their distances from the image centre (normalised between 0 and 1), across the entire dataset. In the low-probability (test) subset, the mean distance is 0.048 and the median distance is 0.022, while on the high-probability (lens) subset, the mean distance is 0.033 and the median distance is 0.022, further underlying the ability of the models to focus on the objects particularly in the case of high-probability lenses.

\begin{table}[h]
    \centering
    \caption{Localization accuracy for the \ac{KiDS} dataset.}
    \label{tab:kids_loc}
    \begin{tabular}{lcc}
        \toprule
        \textbf{Dataset} & \textbf{Bounding Box Size (\%)} & \textbf{Localization Accuracy (\%)} \\
        \midrule
        KiDS (test) & 54.31 & 97.76 \\
        KiDS (lens) & 68.96 & 99.61 \\
        \bottomrule
    \end{tabular}
\end{table}

Table \ref{tab:kids_loc} reports the localization accuracy for the \ac{KiDS} dataset. In this case, bounding boxes are generally bigger than those in \ac{DGL}, with the lens subset having bigger bounding boxes because of the magnification introduced by the candidate lens. Localization accuracy is comparable to that of \ac{DGL} ($> 97\%$), with no significant differences in the two \ac{KiDS} subsets.

\subsection{SDSS}
\label{sec:sdss_results}

\begin{table}[]
\centering
\caption{\ac{MAE} of the models trained on the \ac{SDSS} dataset. The results are taken on the test set and lens set for the models with the combination of hyperparameters with the best performance on the validation set.  All numerical results are rounded to the third decimal place for consistency. The best results, and those falling within their confidence intervals, are highlighted in bold.}
\label{tab:sdss_mae}
\begin{tabular}{@{}cc@{}}
\toprule
 & \textbf{SDSS} \\ \midrule
\textbf{HOG + SVR} & 0.055 $\pm$ 0.000 \\
\textbf{A1} & 0.013 $\pm$ 0.000 \\
\textbf{A3} & 0.012 $\pm$ 0.000 \\
\textbf{NetZ} & 0.012 $\pm$ 0.000 \\
\textbf{PhotoZ} & 0.013 $\pm$ 0.000 \\
\textbf{EfficientNet} & 0.014 $\pm$ 0.000 \\
\textbf{MLPm} & \textbf{0.011 $\pm$ 0.000} \\
\textbf{ResNet} & 0.013 $\pm$ 0.000 \\
\textbf{SwinT} & 0.016 $\pm$ 0.000 \\ \midrule
\textbf{Ensemble} & 0.013 $\pm$ 0.000 \\ \midrule
\textbf{Improvement {[}\%{]}} & 5 \\ \bottomrule
\end{tabular}
\end{table}

\begin{table}[]
\centering
\caption{\ac{NMAD} of the models trained on the \ac{SDSS} dataset. The results are taken on the test set and lens set for the models with the combination of hyperparameters with the best performance on the validation set.  All numerical results are rounded to the third decimal place for consistency. The best results, and those falling within their confidence intervals, are highlighted in bold.}
\label{tab:sdss_nmad}
\begin{tabular}{@{}cc@{}}
\toprule
 & \textbf{SDSS} \\ \midrule
\textbf{HOG + SVR} & 0.060 $\pm$ 0.001 \\
\textbf{A1} & 0.013 $\pm$ 0.000 \\
\textbf{A3} & 0.012 $\pm$ 0.000 \\
\textbf{NetZ} & 0.012 $\pm$ 0.000 \\
\textbf{PhotoZ} & 0.013 $\pm$ 0.000 \\
\textbf{EfficientNet} & 0.014 $\pm$ 0.000 \\
\textbf{MLPm} & \textbf{0.011 $\pm$ 0.000} \\
\textbf{ResNet} & 0.013 $\pm$ 0.000 \\
\textbf{SwinT} & 0.016 $\pm$ 0.000 \\ \midrule
\textbf{Ensemble} & 0.013 $\pm$ 0.000 \\ \midrule
\textbf{Improvement {[}\%{]}} & 5 \\ \bottomrule
\end{tabular}
\end{table}

\begin{table}[]
\centering
\caption{$\sigma_{68}$ of the models trained on the \ac{SDSS} dataset. The results are taken on the test set and lens set for the models with the combination of hyperparameters with the best performance on the validation set.  All numerical results are rounded to the third decimal place for consistency. The best results, and those falling within their confidence intervals, are highlighted in bold.}
\label{tab:sdss_sigma}
\begin{tabular}{@{}cc@{}}
\toprule
 & \textbf{SDSS} \\ \midrule
\textbf{HOG + SVR} & 0.057 $\pm$ 0.000 \\
\textbf{A1} & 0.013 $\pm$ 0.000 \\
\textbf{A3} & \textbf{0.012 $\pm$ 0.000} \\
\textbf{NetZ} & \textbf{0.012 $\pm$ 0.000} \\
\textbf{PhotoZ} & 0.013 $\pm$ 0.000 \\
\textbf{EfficientNet} & 0.014 $\pm$ 0.000 \\
\textbf{MLPm} & \textbf{0.012 $\pm$ 0.000} \\
\textbf{ResNet} & 0.013 $\pm$ 0.000 \\
\textbf{SwinT} & 0.017 $\pm$ 0.000 \\ \midrule
\textbf{Ensemble} & 0.013 $\pm$ 0.000 \\ \midrule
\textbf{Improvement {[}\%{]}} & 4 \\ \bottomrule
\end{tabular}
\end{table}

\begin{table}[]
\centering
\caption{Bias of the models trained on the \ac{SDSS} dataset. The results are taken on the test set and lens set for the models with the combination of hyperparameters with the best performance on the validation set.  All numerical results are rounded to the third decimal place for consistency. The best results, and those falling within their confidence intervals, are highlighted in bold.}
\label{tab:sdss_bias}
\begin{tabular}{@{}cc@{}}
\toprule
 & \textbf{SDSS} \\ \midrule
\textbf{HOG + SVR} & 0.024 $\pm$ 0.001 \\
\textbf{A1} & 0.001 $\pm$ 0.000 \\
\textbf{A3} & \textbf{-0.000 $\pm$ 0.000} \\
\textbf{NetZ} & 0.001 $\pm$ 0.000 \\
\textbf{PhotoZ} & \textbf{0.000 $\pm$ 0.000} \\
\textbf{EfficientNet} & -0.003 $\pm$ 0.000 \\
\textbf{MLPm} & \textbf{0.000 $\pm$ 0.000} \\
\textbf{ResNet} & 0.003 $\pm$ 0.000 \\
\textbf{SwinT} & 0.001 $\pm$ 0.000 \\ \midrule
\textbf{Ensemble} & 0.002 $\pm$ 0.000 \\ \bottomrule
\end{tabular}
\end{table}

\begin{table}[]
\centering
\caption{Outlier rate of the models trained on the \ac{SDSS} dataset. The results are taken on the test set and lens set for the models with the combination of hyperparameters with the best performance on the validation set.  All numerical results are rounded to the third decimal place for consistency. The best results, and those falling within their confidence intervals, are highlighted in bold.}
\label{tab:sdss_outlier}
\begin{tabular}{@{}cc@{}}
\toprule
\textbf{} & \textbf{SDSS} \\ \midrule
\textbf{HOG + SVR} & 0.412 $\pm$ 0.005 \\
\textbf{A1} & 0.008 $\pm$ 0.000 \\
\textbf{A3} & \textbf{0.006 $\pm$ 0.000} \\
\textbf{NetZ} & \textbf{0.008 $\pm$ 0.000} \\
\textbf{PhotoZ} & 0.012 $\pm$ 0.000 \\
\textbf{EfficientNet} & 0.012 $\pm$ 0.000 \\
\textbf{MLPm} & \textbf{0.006 $\pm$ 0.000} \\
\textbf{ResNet} & 0.010 $\pm$ 0.000 \\
\textbf{SwinT} & 0.018 $\pm$ 0.001 \\ \midrule
\textbf{Ensemble} & 0.008 $\pm$ 0.000 \\ \midrule
\textbf{Improvement {[}\%{]}} & 0 \\ \bottomrule
\end{tabular}
\end{table}

Table \ref{tab:sdss_mae} reports the \ac{MAE} on the \ac{SDSS} dataset for all models, using the test set for evaluation and selecting the best-performing hyperparameter configurations from the \ac{KiDS} test set. Across all models, the \ac{MLPm} architecture achieves the best individual performance with an MAE of 0.011, similar to A3 and A1 with scores of 0.012 and 0.013, respectively. \ac{HOG} + \ac{SVR}, a traditional feature-based approach, is less effective than other algorithms and has a \ac{MAE} of 0.055, which is 5 times higher than the best-performing deep learning model.
\ac{LRE} has a \ac{MAE} of 0.013, slightly above the best individual model. The marginal 5\% improvement of \ac{MLPm} in performance over the baseline models (\ac{HOG} + \ac{SVR}, A1 and A3) is relatively modest, indicating that the choice of the architecture is less relevant than in the other presented experiments. Overall, the best results are comparable to those of LSST-wide and DESI-DOT and demonstrate the effectiveness of the proposed deep architectures on datasets with different characteristics.

As observed in the other datasets, the better results of the \ac{DL} algorithms presented in this work are also observed when comparing the $R^2$ scores. In this case, the \ac{DL} architectures used in this work achieve $R^2 \approx 0.87$, compared to $R^2 \approx 0.33$ for baseline architectures. Still, the low average $R^2$ for the baseline architectures is influenced significantly by \ac{HOG} + \ac{SVR} ($R^2 \approx -0.81$). Considering the baseline \ac{DL} architectures, the results are comparable ($R^2 \approx 0.90$) and similar to those of the best model ($R^2 \approx 0.91$).

On this dataset, using \ac{MAE} as both the loss and early stopping criterion yields the best performance, reducing the test \ac{MAE} by approximately 14\% compared to the combination of \ac{MSE} loss and $R^2$ for early stopping.

\autoref{tab:sdss_nmad} and \autoref{tab:sdss_sigma} present the results for the dispersion metrics \ac{NMAD} and $\sigma_{68}$, respectively. Compared to the other datasets, the performance gap between the proposed architectures and the baselines is smaller. \ac{MLPm} achieves the lowest \ac{NMAD} ($\approx 0.011$), with a slight improvement of $\approx 5\%$ over the best baseline. For $\sigma_{68}$, the baseline architectures (e.g., A3 and NetZ) perform comparably to \ac{MLPm}, with values of $\approx 0.012$. These result suggests that for the \ac{SDSS} dataset, the increased complexity of the new architectures does not provide a significant advantage in scatter reduction over standard \ac{DL} baselines. \ac{HOG} + \ac{SVR} remains significantly worse than all \ac{DL} methods, with dispersion values more than 5 times higher than those of the best architecture. A paired bootstrap test (16th–84th percentile confidence interval) on \ac{NMAD} confirms a statistically significant improvement of our best architecture over the baselines.

\autoref{tab:sdss_bias} shows that the majority of the \ac{DL} architectures are  unbiased. \ac{MLPm}, PhotoZ, and A3 achieve a bias of 0.000, while others like EfficientNet show a small bias ($\approx -0.003$). \ac{HOG} + \ac{SVR} has a higher bias ($\approx 0.024$), further confirming its limitations compared to deep learning approaches.

Finally, \autoref{tab:sdss_outlier} highlights the performance of \ac{DL} models in rejecting outliers on this dataset. Most architectures, including the baselines (A1, A3, NetZ, PhotoZ) and the proposed \ac{MLPm}, EfficientNet, and ensembling, achieve an outlier rate $< 1\%$, with A3 and \ac{MLPm} being the best ones. Consequently, no significant improvement is observed, as both the baseline and advanced models effectively remove outliers. This result further indicates that the \ac{SDSS} test set presents fewer challenges for these models compared to datasets like DES-deep or \ac{KiDS}.

Training the optimal \ac{MLPm} architecture on larger ($90\times90$ pixel) images without resampling produced results comparable to NetZ and A3, albeit with higher errors than in the $45\times45$ pixel case.

Finally, the work in \cite{dey_photometric_2022} reports competitive performance on \ac{SDSS} using a deep capsule network and evaluates the results using $\sigma_{MAD}$, defined as:

\begin{equation}
    \sigma_{MAD} = \text{median}\left( \left| \frac{\hat{y} - y}{1 + y} - \text{median}\left( \frac{\hat{y} - y}{1 + y} \right) \right| \right)
\end{equation}

They achieve $\sigma_{MAD} \approx 0.009$ using also auxiliary morphological class labels during training. In this work, we obtain better results ($\sigma_{MAD} \approx 0.008$) with \ac{MLPm} and \ac{ResNet}, which use only raw image data.

\subsubsection{Explainability}

\begin{figure}
    \centering
    \includegraphics[width=\linewidth]{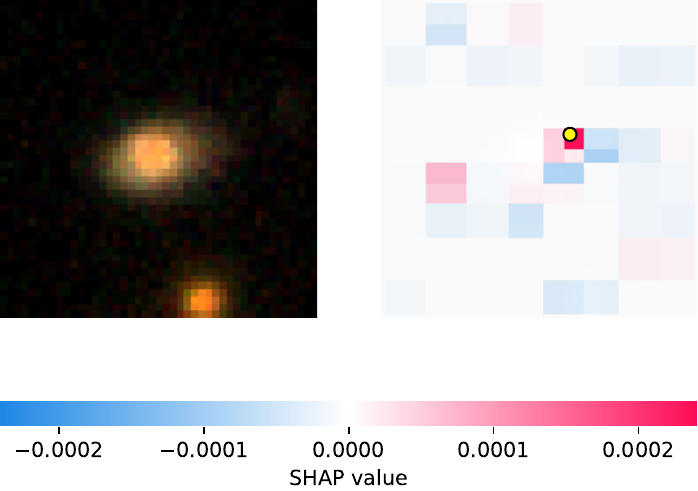}
    \caption{Example of a \ac{SHAP} heatmap on the \ac{SDSS} dataset. 
    The yellow dot indicates the position of the most influential pixel according to \ac{SHAP}. Each side corresponds to $\approx 18$ arcsec.}
    \label{fig:sdss_shap_98}
\end{figure}

\begin{figure}
    \centering
    \includegraphics[width=\linewidth]{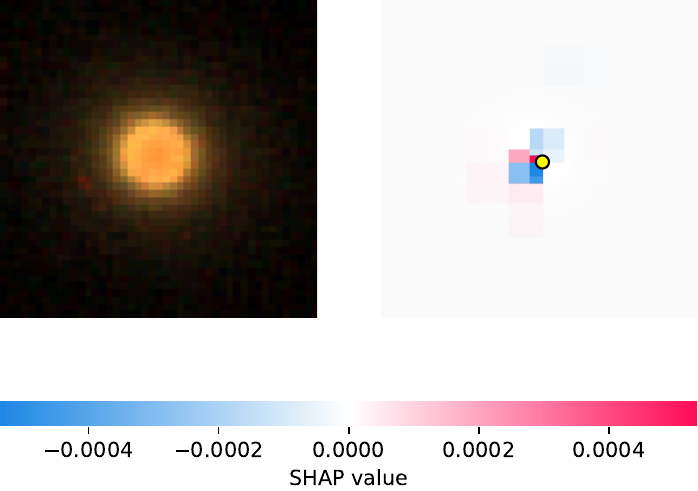}
    \caption{Example of a \ac{SHAP} heatmap on the \ac{SDSS} dataset. 
    The yellow dot indicates the position of the most influential pixel according to \ac{SHAP}. Each side corresponds to $\approx 18$ arcsec.}
    \label{fig:sdss_shap_14}
\end{figure}

\begin{figure}
    \centering
    \includegraphics[width=\linewidth]{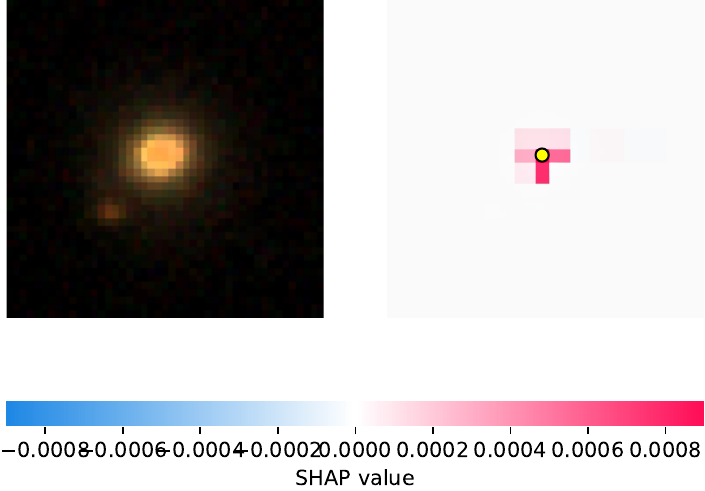}
    \caption{Example of a \ac{SHAP} heatmap on the \ac{SDSS} dataset. 
    The yellow dot indicates the position of the most influential pixel according to \ac{SHAP}. Each side corresponds to $\approx 18$ arcsec.}
    \label{fig:sdss_shap_19}
\end{figure}

\begin{figure}
    \centering
    \includegraphics[width=\linewidth]{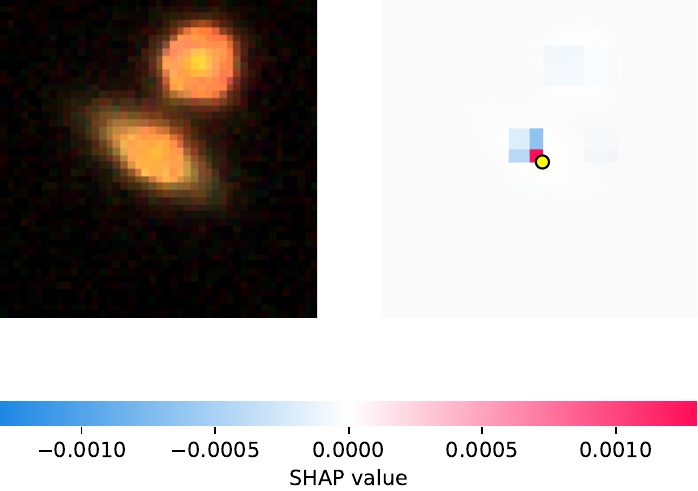}
    \caption{Example of a \ac{SHAP} heatmap on the \ac{SDSS} dataset. 
    The yellow dot indicates the position of the most influential pixel according to \ac{SHAP}. Each side corresponds to $\approx 18$ arcsec.}
    \label{fig:sdss_shap_11}
\end{figure}

Figures \ref{fig:sdss_shap_98}, \ref{fig:sdss_shap_14}, \ref{fig:sdss_shap_19}, 
and \ref{fig:sdss_shap_11} show representative examples of \ac{SHAP} heatmaps for 
\ac{SDSS} galaxy redshift estimation obtained from the best-performing model. The model's focus changes based on the image content. \autoref{fig:sdss_shap_98} shows the case of a diffuse heatmap caused by the presence of two sources, with the most influential pixel not aligned with the center of the central galaxy (the object of interest). \autoref{fig:sdss_shap_14} and \autoref{fig:sdss_shap_19}, on the other hand, show that \ac{SHAP} focuses only on the central parts of the images, as the sources are more compact, similar to the case of DESI-DOT. Finally, \autoref{fig:sdss_shap_11} shows that the model focuses on the central part of the image, even if there are two bright bodies. 
In this case, the most influential pixel is correctly aligned with the center of the object of interest.

\begin{figure}
    \centering
    \includegraphics[width=0.8\linewidth]{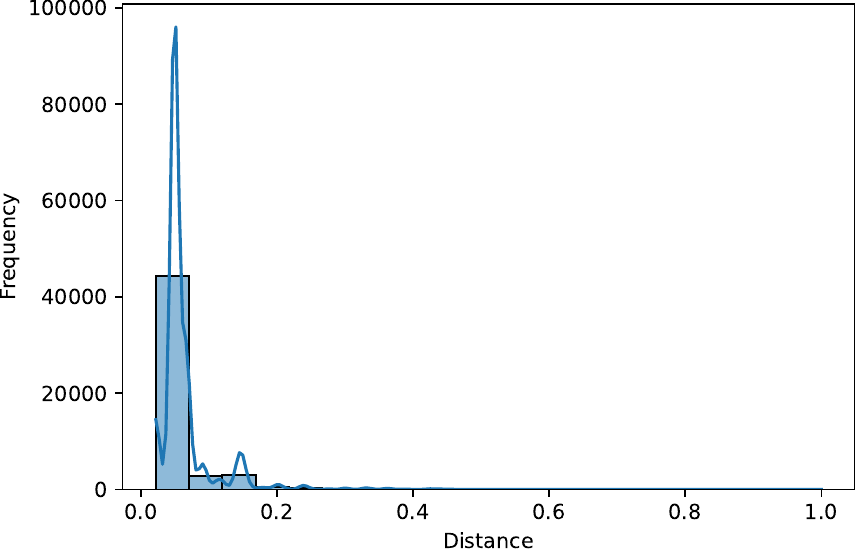}
    \caption{Distribution of the distances between the most influential pixels and the centres of the images for the \ac{SDSS} dataset. The distances range between 0 and 1, where 1 indicates the maximum possible distance in a single image.}
    \label{fig:sdss_shap_distributions}
\end{figure}

To quantify the relevance of the most influential pixels, \autoref{fig:sdss_shap_distributions} shows the distribution of their distances from the image centre (normalised between 0 and 1), across the entire dataset. Overall, the mean distance is 0.065 and the median distance is 0.050, indicating the ability of the best model to focus on the galaxies for which redshift is estimated.

In this dataset, the average bounding box size is $\approx 58.94\%$ of the image, similar to \ac{KiDS}, and the localization accuracy is $99.69\%$, comparable to the other datasets (except DES-deep).

\subsection{Computational cost}

\begin{table}[ht]
    \centering
    \caption{Computational cost for training the proposed architectures on all the datasets}
    \label{tab:computational_cost}
    \begin{tabular}{llrrr}
        \toprule
        \textbf{Model} & \textbf{Training time} & \textbf{\ac{DGL}} & \textbf{\ac{KiDS}} & \textbf{\ac{SDSS}} \\
        \midrule
        \textbf{\ac{SwinT}} & Epoch [s] & 18 & 7 & 480 \\
                   & Total [h] & 3 & 1 & 13 \\
                   & Hyperparameters tuning [h] & 486 & 162 & 13 \\
        \addlinespace
        \textbf{\ac{MLPm}} & Epoch [s] & 24 & 8 & 840 \\
                  & Total [h] & 3 & 1 & 23 \\
                  & Hyperparameters tuning [h] & 324 & 108 & 23 \\
        \addlinespace
        \textbf{ResNet} & Epoch [s] & 14 & 4 & 360 \\
               & Total [h] & 2 & 1 & 10 \\
               & Hyperparameters tuning [h] & 108 & 54 & 10 \\
        \addlinespace
        \textbf{EfficientNet} & Epoch [s] & 26 & 12 & 1440 \\
                     & Total [h] & 4 & 2 & 40 \\
                     & Hyperparameters tuning [h] & 216 & 108 & 40 \\
        \midrule
        \textbf{Ensemble} & Total [h] & 11 & 4 & 87 \\
        & Hyperparameters tuning [h] & 1134 & 432 & 87 \\
        \bottomrule
    \end{tabular}
\end{table}

This section analyzes the computational cost for training the proposed architectures on different datasets. Since the four \ac{DGL} datasets have the same characteristics (image size and number of samples), they are analyzed together. All the architectures were trained on a single NVIDIA Ampere A100 GPU. Table~\ref{tab:computational_cost} summarizes the training duration per epoch (in seconds), the total training time (in hours) for each architecture type across \ac{DGL}, \ac{KiDS}, and \ac{SDSS} and the cost of hyperparameter tuning. For each architecture, the reported training time refers to the hyperparameters leading to the highest parameters number and assumes no early stopping to provide a conservative estimate. The computational cost varies significantly depending on the architecture. ResNet is the most efficient architecture, requiring only 14 seconds per epoch on \ac{DGL} and maintaining the lowest total training time across all scenarios (e.g., 10 hours for \ac{SDSS}). EfficientNet is the most computationally expensive (e.g., 40 hours total for \ac{SDSS}). The computational cost depends also by the dimension of the training set. Both \ac{DGL} and \ac{KiDS} are relatively small, allowing for rapid training (up to 4 hours per architecture). The \ac{SDSS} dataset is characterized by a larger number of training samples and highlights the scaling challenge posed by large-scale astronomical surveys.

\subsection{Threats to validity}
\label{sec:threats_validity}

Considering simulated gravitational lenses, real lens candidates, and non-lensed galaxies introduces threats to the validity of the proposed image-only pipeline. We identify three challenges regarding the interpretation and generalization of our results:

\begin{itemize}
    \item Heterogeneous \ac{GT} definitions: the definition of the target variable varies across datasets. While \ac{SDSS} relies on spectroscopic redshifts, \ac{KiDS} utilizes photometric estimates. Consequently, in the case of \ac{KiDS}, the models approximate the output of a higher-dimensional estimator rather than predicting physical redshift directly.
    
    \item Morphological and \ac{PSF} differences: the  features correlating with redshift in Einstein rings can differ from those in standard spiral or elliptical galaxies. In addition, the difference between the idealized \acp{PSF} in simulations and the noise artifacts in real surveys challenges the models' ability to generalize. Still, this choice is unavoidable because of the limited number of confirmed gravitational lenses and gravitationally-lensed supernovae.
    
    \item Unimodality limitations: by restricting the input to images, the pipeline neglects spectral features and additional multimodal information. The  proposed approach relies on the ability of deep learning architectures to identify morphological and color-based features that correlate with redshift.
\end{itemize}

\section{Conclusions and future work}
\label{sec:conclu}

This work focuses on redshift estimation using multi-channel simulated and real images of gravitationally-lensed supernovae, gravitational lenses, and galaxies. We propose \arch, a deep learning pipeline, based on computer vision models belonging to different families. More in detail, \ac{ResNet} and EfficientNet are \acp{CNN}, \ac{MLPm} relies on \acp{MLP}, and \ac{SwinT} is attention-based.
The results demonstrate the effectiveness of the models on datasets with different characteristics and the potential for accurately estimating redshift in both existing and upcoming real data. The results also highlight that all the architectures are less effective on noisy datasets. Overall, most of the deep learning models exhibit high $R^2$ scores ($> 0.9$) on most of the simulated datasets. 

The results of \ac{HOG} + \ac{SVR} suggest that shape-based features capture only some information about redshift. Other existing architectures (A1, A3, PhotoZ and NetZ) have slightly better performance, but \arch \ac{DL} architectures, based on convolutional and non-convolutional neural networks, and their ensembles, are better, on average, at redshift estimation.

On the \ac{DGL} simulated datasets, EfficientNet is the best network on average, being particularly effective for DES-deep, a noisy dataset. In the other simulated datasets, most of the models exhibit similar behaviour ($R^2 > 0.9$). Ensembling the outputs of multiple networks further improves the results.

For the \ac{KiDS} dataset, the best models are \ac{MLPm} and \ac{SwinT}. They learn to estimate redshift from low-probability lens candidates and are able to generalize to high-probability lens candidates.

The \ac{SDSS} dataset, different from \ac{DGL} and \ac{KiDS}, includes only non-lensed galaxies. Notably, the hyperparameters and models optimized on \ac{KiDS} transfer effectively to \ac{SDSS}, indicating good generalization. The architectures proposed in \arch prove robust for redshift estimation, outperforming the strongest baselines (A3 and NetZ) by $\approx 5\%$ in terms of \ac{MAE} and \ac{NMAD}.

The analysis of the models' explainability using \ac{SHAP} shows that they give more importance to the central regions of the images on both the simulated and real datasets. This result is quantified using localization accuracy, higher than 95\% on all the datasets except DES-deep. This dataset contains gravitational lenses that spread through the entire image and the images have a lower quality, which poses challenges also in terms of classification \cite{PinciroliVago2023}.

Future work can explore related research areas, including:

\begin{itemize}
    \item Developing models that generalize across datasets and can be applied to data from different instruments.

    \item Incorporating time-series into the input data, similar to \ac{DGL} \cite{PinciroliVago2023}, that classifies multimodal observations consisting of images and time series, and Hydra \cite{pinciroli_vago_multimodal_2025}, that counts objects in the same multimodal data.

    \item Validating models on upcoming real observations, such as the ones from LSST, to identify new gravitationally-lensed supernovae, as proposed in \cite{PinciroliVago2023}, count objects in multimodal data \cite{pinciroli_vago_multimodal_2025}, and estimate redshift.
\end{itemize}

To conclude, the proposed research directions serve as a foundation for applying interpretable \ac{DL} algorithms to upcoming large sky surveys of astronomical data, aiming to estimate redshift accurately from images and multimodal data.

\section*{Acknowledgments}
NOPV acknowledges support from INAF and CINECA for granting 125,000 core hours on the Leonardo supercomputer. We acknowledge Petrillo et al. (2019, MNRAS, 484, 3879), Li et al. (2020, ApJ, 899, L30) and Li et al. (2021, APJ, 923, 16). Based on observations made with ESO Telescopes at the La Silla Paranal Observatory under programme IDs 177.A-3016, 177.A-3017, 177.A-3018 and 179.A-2004, and on data products produced by the KiDS consortium. The KiDS production team acknowledges support from: Deutsche Forschungsgemeinschaft, ERC, NOVA and NWO-M grants; Target; the University of Padova, and the University Federico II (Naples). We acknowledge Kuijken et al. (2019, A\&A 625, A2).

\clearpage

\bibliographystyle{unsrt}  
\bibliography{sn-bibliography}

\end{document}